\documentclass[lettersize,journal]{IEEEtran}
\usepackage{amsmath,amsfonts}
\usepackage{algorithmic}
\usepackage{algorithm}
\usepackage{array}
\usepackage[caption=false,font=normalsize,labelfont=sf,textfont=sf]{subfig}
\usepackage{textcomp}
\usepackage{stfloats}
\usepackage{url}
\usepackage{verbatim}
\usepackage{graphicx}
\usepackage{hyperref}
\usepackage{color}
\usepackage{bm}
\usepackage{newtxtext,newtxmath}
\usepackage{subfig}
\usepackage{multirow}
\usepackage{diagbox}
\usepackage{xcolor}

\usepackage{etoolbox}
\makeatletter
\def\@IEEEBIOskipN{2\baselineskip}
\expandafter\patchcmd\csname\string\IEEEbiography\endcsname
  {\vskip \@IEEEBIOskipN plus 1fil minus 0\baselineskip}
  {\vskip \@IEEEBIOskipN}{}{\typeout{** BIO PATCH FAILED}}
\patchcmd{\IEEEbiographynophoto}
  {\vskip 4\baselineskip plus 1fil minus 0\baselineskip}
  {\vskip \@IEEEBIOskipN}{}{\typeout{** BIO PATCH FAILED}}
\makeatother

\begin{document}

\title{Deep Learning-Enhanced Real-Time Wi-Fi Sensing Through Single Transceiver Pair}

\author{Yuxuan~Liu, 
        Chiya~Zhang,
        Yifeng~Yuan,
        Chunlong~He,
        Weizheng~Zhang,
        Gaojie~Chen
        \thanks{This work was supported by the National Natural Science Foundation of China under Grants 62394294 and 62394290, and the Shenzhen Key Industrial Research and Development Project under Grant ZDCY20250901111405006. It was also supported by the Foundation of the National Key Laboratory of Radar Signal Processing under Grant JKW202303, and the Key Field Special Project for Guangdong Provincial Ordinary Universities under Grant 2024ZDZX1016. This work was supported in part by the Fundamental and Interdisciplinary Disciplines Breakthrough Plan of the Ministry of Education of China under Grant JYB2025XDXM406, and by the Fundamental Research Funds for the Central Universities, Sun Yat-sen University, under Grant No. 24hytd010. }
        \thanks{Y. Liu, Y. Yuan and W. Zhang are with the School of Electronic and Information Engineering, Harbin Institute of Technology, Shenzhen 518055, China.}
        \thanks{C. Zhang is with the School of Electronic and Information Engineering, Harbin Institute of Technology, Shenzhen 518055, China, and also with the Peng Cheng Laboratory (PCL), Shenzhen 518055, China.}
        \thanks{C. He is with the Guangdong Key Laboratory of Intelligent Information Processing, Shenzhen University, Shenzhen 518060, China.}
        \thanks{Gaojie Chen is with the School of Flexible Electronics (SoFE), Sun Yat-sen University, Shenzhen, Guangdong 518107, China, and also with the State Key Laboratory of Flexible Electronics (LoFE), Nanjing University of Posts \& Telecommunications, 9 Wenyuan Road, Nanjing 210023, China. (e-mail: chengj235@mail.sysu.edu.cn). }
        \thanks{The Corresponding Author is Chiya Zhang (email:zhangchiya@hit.edu.cn).}
        }

\IEEEpubid{}

\maketitle

\IEEEpubidadjcol

\begin{abstract}

The advancement of next-generation Wi-Fi technology heavily relies on sensing capabilities, which play a pivotal role in enabling sophisticated applications. In response to the growing demand for large-scale deployments, contemporary Wi-Fi sensing systems strive to achieve high-precision perception while maintaining minimal bandwidth consumption and antenna count requirements. Remarkably, various deep learning-driven perception technologies have demonstrated the ability to surpass conventional resolution limits. However, the theoretical underpinnings of this phenomenon have not been thoroughly investigated in existing research. We find that under hardware-constrained conditions, the performance gains of deep learning in Wi-Fi sensing primarily originate from two aspects: prior information and temporal correlation, which act as specific forms of side information that reduce the estimation error bound. We construct a deep learning-based Wi-Fi sensing system using only a single transceiver pair and design experiments to validate these gains. The system achieves an average human pose estimation error of 0.2189 m and an average localization error of 0.6124 m, while operating in real time at 42 fps on commodity hardware.
\end{abstract}


\section{Introduction}
\IEEEPARstart{S}{ensing} plays a crucial role in the evolution of next-generation Wi-Fi technology \cite{reshef2022future}. With its high penetration rate and diverse non-contact sensing functionalities, device-free Wi-Fi sensing has presented extensive application potential in areas such as home security, abnormal behavior detection, and elderly or child care \cite{ahmad2024wifi}. 

Despite the significant potential of device-free Wi-Fi sensing technologies, their large-scale implementation remains challenging. The received signal strength indicator (RSSI) \cite{RSSI, bao2025rssi} or channel state information (CSI) \cite{li2022toward} of a target at different positions in the environment can be used as fingerprints. Researchers construct a fingerprint database for indexing and compare received fingerprint data with the stored entries. The spatial coordinates corresponding to the closest match in the database are then used as the estimated positioning result. These methods can only be used in locations where large amounts of data have been collected in advance. Another approach employs radar technology for sensing, with inverse synthetic aperture radar (ISAR) \cite{WiFi_wall} and multiple signal classification (MUSIC) \cite{kotaru2015spotfi} along with its derived algorithms \cite{li2016dynamic} being representative methods in this category. These technologies often require a large number of antennas and wide bandwidth that are difficult to implement with commodity Wi-Fi devices.

Some studies have proposed device-free Wi-Fi sensing technology suitable for large-scale applications. For instance, Widar2.0 \cite{qian2018widar2} requires at least one transmitting antenna and three receiving antennas to facilitate positioning; however, its limited system resolution hampers more precise sensing applications such as human pose estimation (HPE) or breathing detection.

Recently, methods leveraging deep learning have garnered increasing attention, demonstrating superior resolution and accuracy compared to traditional technologies. Most of these methods leverage multiple transceiver units to enhance accuracy \cite{jeong2025ubigest, yan2024person}, while others focus on achieving high-precision sensing with minimal hardware \cite{li2020wihf}, demonstrating notable success. For instance, CSI2Depth \cite{alvarez2025csi2depth} utilizes the Wi-Fi CSI data from the MM-Fi dataset \cite{yang2023mm}, generating depth images based on CSI collected by a pair of TP-Link N750 Wi-Fi APs. 

Notably, some deep learning-based methods have demonstrated precision that significantly exceeds the theoretical resolution limit in radar theory (see Section II). Although certain studies suggest that the Fresnel zone model \cite{ma2019wifi} can explain the high accuracy of Wi-Fi sensing, when only a single transmitter-receiver pair is used, this model only accounts for the system’s high precision in the direction perpendicular to the line connecting the transceivers. According to this model, additional devices must be deployed to achieve similar accuracy in other directions. The model in \cite{wang2015understanding} reveals the influence of target motion on CSI amplitude, providing a theoretical basis for motion pattern recognition based on CSI. However, it still fails to explain why deep learning-based methods also achieve remarkable performance in regression tasks such as HPE. To the best of our knowledge, there is currently no widely accepted theoretical framework that can fully explain where the performance gains brought by deep learning to Wi-Fi sensing originate.

In this paper, we first propose a theoretical explanation for the performance gains achieved by deep learning. Then, based on our theoretical insight, we develop a deep learning-based real-time Wi-Fi sensing and visualization system and conduct experiments to validate the proposed theoretical explanation. The main contributions of this work are as follows:
\begin{itemize}
    \item We establish a theoretical framework based on information theory to explain the performance gains of deep learning in Wi‑Fi sensing. Through Theorem 1, we show that side information reduces the MSE lower bound by a factor of $e^{-\frac{2}{p} I(\bm{X};\bm{Y}\vert\bm{Z})}$ and identify two specific forms of side information in this context, prior information about the target structure and temporal correlation across sequential measurements, which enable deep learning to surpass conventional sensing capability limits.
    \item We construct a deep learning-based Wi‑Fi sensing system using only a single transceiver pair, and experimentally validate both identified sources of gain. Prior information is shown to be critical for effective model training, while temporal correlation is demonstrated to directly improve sensing accuracy over time. The system achieves an average HPE error of $0.2189\ \text{m}$ and an average localization error of $0.6124\ \text{m}$.
    \item We implement a real-time sensing and visualization pipeline on commodity hardware, achieving a sustained frame rate of $42\ \text{fps}$ through parallel data processing and event-driven file monitoring.
\end{itemize}

The remainder of this paper is organized as follows. Section II presents the theoretical analysis of the performance gains achieved through deep learning. Section III introduces the design and implementation of the proposed real-time Wi-Fi sensing and visualization system. Section IV details the experimental setup and results. Finally, Section V concludes the paper.

\section{Theoretical Analysis of the Performance Gains}

\subsection{Conventional Sensing Capability Analysis}

A single-transceiver-pair Wi-Fi system can be analyzed from two complementary perspectives: the Rayleigh resolution criterion and the Fresnel zone model. Both converge on the same conclusion: under these hardware constraints, the system cannot achieve the uniform high precision required for tasks such as HPE. 

The Rayleigh resolution criterion analyzes this limitation from two aspects: angular resolution and distance resolution. We consider a general scenario in which the distance between the transmitter and receiver is arbitrary. 

\begin{figure}[ht]
    \centering 
    \includegraphics[width=0.8\columnwidth]{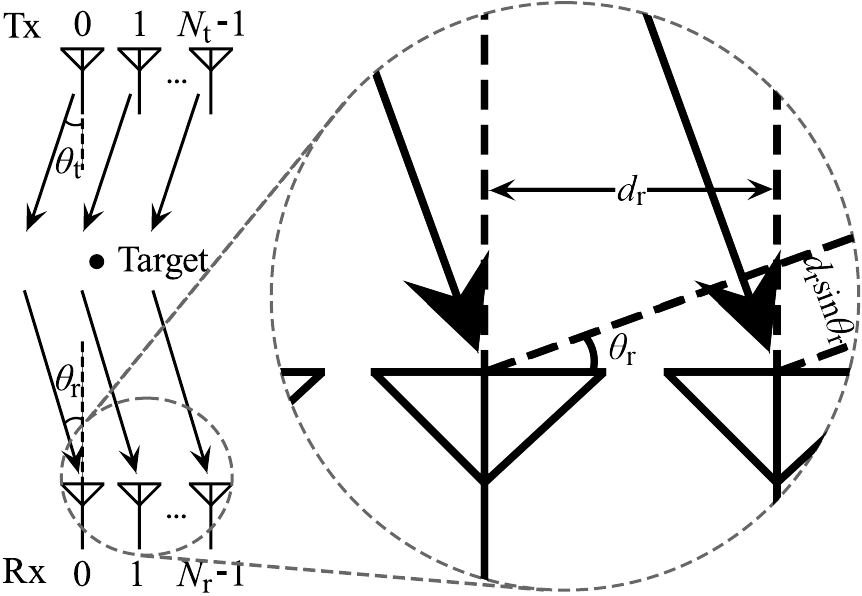}
    \caption{Angle-of-arrival estimation in a typical Wi-Fi sensing system. The angle of arrival is determined from the time difference between the arrivals of the same signal at different antennas.}
    \label{radar_system}
\end{figure}

For angle-of-arrival (AoA) estimation, consider a signal transmitted from the transmitting antennas, reflected by a target, and finally received by the receiving antennas, as shown in Fig. \ref{radar_system}, where the angle-of-departure is $\theta_{\text{t}}$, and the AoA is $\theta_{\text{r}}$. The distance between adjacent transmitting antennas is denoted by $d_{\text{t}}$, and the distance between adjacent receiving antennas is denoted by $d_{\text{r}}$. Denote the signal from transmitting antenna $m$ to the receiving antenna $n$ as $\text{signal}_{mn}$, then the phase difference between $\text{signal}_{mn}$ and $\text{signal}_{00}$ is
\begin{equation}
    \label{radar_phase_diff}
    \Delta \phi_{mn} = \frac{2 \pi}{\lambda}(m d_{\text{t}} \sin \theta_{\text{t}} + n d_{\text{r}} \sin \theta_{\text{r}}),
\end{equation}
where $\lambda$ is the wavelength of the transmitted signal. Let the transmitted signal be $x(t)$; then the received signal can be represented as
\begin{equation}
    \label{radar_recv}
    y(t) = \sum_{m = 0}^{m = N_{\text{t}}-1} \sum_{n = 0}^{n = N_{\text{r}}-1} x(t) e^{j \Delta \phi_{mn}},
\end{equation}
which can be regarded as a 2-D Fourier transform, so the angular resolution of the established system can be regarded as the resolution of this transform in the direction of $\theta_{\text{r}}$. Thus, the angular estimation resolution of the system is
\begin{equation}
    \label{radar_ang_res}
    \Delta \theta_{\text{r}} = \frac{\lambda}{N_{\text{r}} d_{\text{r}} \left\lvert \cos \theta_{\text{r}}\right\rvert }.
\end{equation}
Under typical Wi-Fi sensing system configurations, $d_{\text{r}} = \frac{\lambda}{2}$ and $N_{\text{r}} = 3$, and the resulting angular resolution is $\Delta \theta_{\text{r}} = 0.667 \ \text{rad}$. 

For distance estimation, time-of-flight (ToF) estimation can only determine an ellipse with the transmitter and receiver as its foci, within which the target may be located. Distance estimation from the target to the receiver must therefore be combined with AoA estimation results. Since
\begin{equation}
    \label{diff_dist_cos}
    \text{d}l_\text{t} = \frac{l_{\text{r}} - d_{\text{rt}} \cos \theta_{\text{r}}}{l_{\text{t}}}\text{d}l_{\text{r}},
\end{equation}
where $l_\text{t}$ is the distance between the target and the transmitting antennas, $l_\text{r}$ is the distance between the target and the receiving antennas, $d_{\text{rt}}$ is the distance between the transmitting and receiving antennas. Let $l_\text{t} + l_\text{r} = L$ with $L > d_{\text{rt}}$; then
\begin{equation}
    \label{diff_dist_cos_2}
    \text{d}l_\text{t} = \left(1 - \frac{2 d_{\text{rt}}^2 \sin^2 \theta_{\text{r}}}{d_{\text{rt}}^2 + L^2 - 2 d_{\text{rt}} L \cos \theta_{\text{r}}}\right) \text{d}l_{\text{r}}.
\end{equation}
According to the Rayleigh resolution criterion, $L$ is estimated to have an upper resolution of 
\begin{equation}
    \label{radar_L_res}
    \Delta L = \frac{c}{B},
\end{equation}
where $c$ is the speed of light, and $B$ is the signal bandwidth. Since $\Delta l_{\text{t}} = \Delta L - \Delta l_{\text{r}}$, substituting $\Delta l_{\text{t}}$ and $\Delta l_{\text{r}}$ for $\text{d} l_{\text{t}}$ and $\text{d} l_{\text{r}}$, (\ref{diff_dist_cos_2}) can be transformed into
\begin{equation}
    \label{radar_l_r_res}
    \Delta l_{\text{r}} = \frac{d_{\text{rt}}^2 + L^2 - 2 d_{\text{rt}} L \cos \theta_{\text{r}}}{2 (L - d_{\text{rt}} \cos \theta_{\text{r}})^2} \frac{c}{B}.
\end{equation}

When $\theta_{\text{r}} = 0$, $\Delta l_{\text{r}}$ reaches its minimum value. For typical Wi-Fi sensing systems, the bandwidth ranges from $10 \sim 40 \ \text{MHz}$ \cite{xie2015precise}. Therefore, from (\ref{radar_l_r_res}), the upper distance estimation resolution is $\Delta l_{\text{r}} = 3.75 \ \text{m}$. Then the total spatial resolution is 
\begin{equation}
    \label{radar_res}
    \Delta l = \sqrt{ \Delta l_{\text{r}}^2 + \left(2l_{\text{r}}\tan\frac{\Delta\theta_{\text{r}}}{2}\right)^2}.
\end{equation}
Since $\Delta l \geq \Delta l_{\text{r}}$, the resolution is not enough for accurate sensing tasks such as HPE.

A complementary perspective from physical optics---the Fresnel zone model---further reveals the spatial sensitivity distribution, leading to the same conclusion.

In the Fresnel zone model, for a transmitter and receiver separated by $d_{\text{rt}}$, the $n$-th Fresnel zone boundary satisfies
\begin{equation}
    \label{fresnel_cond}
    l_{\text{t}} + l_{\text{r}} = d_{\text{rt}} + n\frac{\lambda}{2},
\end{equation}
forming concentric ellipsoidal regions with the transmitter and receiver as foci. Crossing a zone boundary changes the phase by $\pi$, producing a significant CSI variation. However, when the target does not cross the region boundary, e.g., when the target moves along the elliptical boundary given by (\ref{fresnel_cond}), the model implies that the target's motion cannot be distinguished. Therefore, according to the Fresnel zone model, the system lacks the ability to resolve the target when it does not cross the Fresnel zone boundary.

The analyses above demonstrate that, from either perspective, a single-transceiver-pair Wi-Fi sensing system cannot provide the uniform high precision required for tasks such as HPE.

\subsection{Performance Gain Brought by Side Information}

We believe that the key to deep learning's remarkable success lies in its ability to effectively leverage side information beyond the current input data. For any estimation problem, especially for complex problems that require deep learning, it is often difficult to model the estimation process. Therefore, it is hard to analyze the impact of side information from the perspective of estimation theory. However, we can analyze the impact of side information from the perspective of information theory.

\begin{figure}[ht]
    \centering                                    
    \includegraphics[width=0.9\columnwidth]{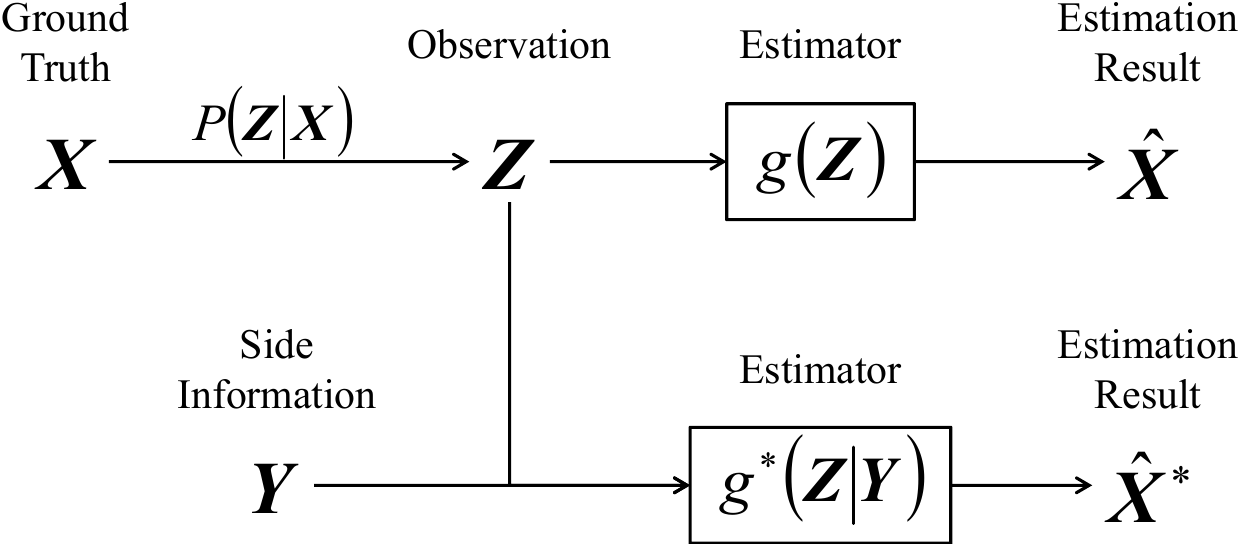}
    \caption{A flowchart of an estimation process, where $g(Z)$ is an estimator that relies only on observations, and $g^\ast (Z|Y)$ is an estimator when side information $Y$ is known.}
    \label{fig_system_model} 
\end{figure}

Consider the estimation process shown in Fig. \ref{fig_system_model}, where $\bm{X} \in \mathbb{R}^p$ is the continuous random vector to be estimated, $\bm{Z}$ is the observation, $\bm{Z} \sim P(\cdot \vert \bm{X})$, $g(\bm{Z})$ is an estimator that relies only on observations, and $\hat{\bm{X}}$ is the corresponding estimation result. $\bm{Y}$ is the side information related to $\bm{X}$, $g^\ast(\bm{Z} \vert \bm{Y})$ is an estimator when the side information $\bm{Y}$ is known, and $\hat{\bm{X}}^\ast$ is its estimation result. Note that in the figure, $\bm{X}$, $\bm{Y}$, $\bm{Z}$, $\hat{\bm{X}}$, and $\hat{\bm{X}}^\ast$ are all random variables, and their samples are denoted by lowercase bold letters. For example, a sample of $\bm{X}$ is denoted by $\bm{x}$. 

Next, we analyze the error of this process, focusing primarily on the mean square error (MSE). This is because, on the one hand, it has many favorable mathematical properties, and on the other hand, for estimation problems using deep learning, when the loss function is set to the MSE, the deep neural network can approximate the minimum mean square error (MMSE) estimator for that problem. 

For any estimator, the lower bound of its MSE is given by the following lemma.

\textbf{Lemma 1.} \emph{For any continuous random vector $\bm{X}$, given observation $\bm{Z}$ and the corresponding estimation result $\hat{\bm{X}}$,}
\begin{equation}
    \label{lemma_1}
    \mathbb{E}\left[\left\lVert \bm{X} - \hat{\bm{X}}\right\rVert ^2\right] \geq \frac{p}{2\pi e} e^{\frac{2}{p} h(\bm{X} \vert \bm{Z})},
\end{equation}
\emph{where $\mathbb{E}\left[\cdot\right] $ represents the expectation, $\left\lVert \cdot \right\rVert^2$ represents the squared Euclidean norm, $p$ is the dimension of $\bm{X}$, and $h(\bm{X} \vert \bm{Z})$ denotes the conditional differential entropy. The inequality holds with equality if and only if the estimator is unbiased, $\bm{X} \vert \bm{Z}$ is Gaussian, and $\bm{\varSigma}\left(\bm{X} \vert \bm{Z}=\bm{z}\right) = \sigma^2 \bm{I}_p$, where $\sigma$ is a constant independent of $\bm{z}$, $\bm{\varSigma}(\cdot)$ denotes the covariance matrix, and $\bm{I}_p$ denotes the $p$-dimensional identity matrix.}

Lemma 1 is the matrix form of Theorem 8.6.6 in \cite{cover2005differential}; a brief proof is provided here.

\emph{Proof.}
\begin{equation}
    \label{proof_lemma_1}
    \begin{aligned}
        \mathbb{E} \left[\left\lVert \bm{X} - \hat{\bm{X}}\right\rVert ^2\right] \geq & \min_{\hat{\bm{X}}} \mathbb{E}\left[\left\lVert \bm{X} - \hat{\bm{X}}\right\rVert ^2\right], \\
        = & \mathbb{E} \left[\left\lVert \bm{X} - \mathbb{E}\left[\bm{X}\right]\right\rVert^2\right], \\
        = & \mathbb{E}_{\bm{Z}} \left[ \text{tr} \left(\bm{\varSigma}\left(\bm{X} \vert \bm{Z}=\bm{z}\right) \right)\right], 
    \end{aligned}
\end{equation}
where $\mathbb{E}_{\bm{Z}}\left[\cdot\right]$ denotes the expectation over different realizations of $\bm{Z}$, $\text{tr}(\cdot)$ denotes the trace of a matrix. Given the covariance matrix $\bm{\varSigma}\left(\bm{X} \vert \bm{Z}=\bm{z}\right)$, the differential entropy $h(\bm{X} \vert \bm{Z}=\bm{z})$ attains its maximum when $\bm{X} \vert \bm{Z}=\bm{z}$ follows a multivariate Gaussian distribution, i.e., 
\begin{equation}
    \label{proof_lemma_2}
    h(\bm{X} \vert \bm{Z}=\bm{z}) \leq \frac{1}{2} \ln \left(\left(2 \pi e\right)^p \det \left(\bm{\varSigma}\left(\bm{X} \vert \bm{Z}=\bm{z}\right)\right) \right) ,
\end{equation}
where $\text{det}\left(\cdot\right)$ represents the determinant of a matrix. Note that
\begin{equation}
    \label{proof_lemma_3}
    \frac{1}{p} \text{tr} \left(\bm{\varSigma}\left(\bm{X} \vert \bm{Z}=\bm{z}\right)\right) \geq \left(\det \left(\bm{\varSigma}\left(\bm{X} \vert \bm{Z}=\bm{z}\right)\right)\right)^{\frac{1}{p}}.
\end{equation}
The equality in (\ref{proof_lemma_3}) holds if and only if all eigenvalues of $\bm{\varSigma}\left(\bm{X} \vert \bm{Z}=\bm{z}\right)$ are equal. Then from (\ref{proof_lemma_2}) and (\ref{proof_lemma_3}), 
\begin{equation}
    \label{proof_lemma_4}
    \mathbb{E}_{\bm{Z}}\left[\text{tr} \left(\bm{\varSigma}\left(\bm{X} \vert \bm{Z}=\bm{z}\right)\right)\right] \geq \mathbb{E}_{\bm{Z}}\left[\frac{p}{2 \pi e} e^{\frac{2}{p} h(\bm{X} \vert \bm{Z}=\bm{z})}\right] .
\end{equation}
By Jensen's inequality, we have
\begin{equation}
    \label{proof_lemma_5}
    \mathbb{E}_{\bm{Z}}\left[\frac{p}{2 \pi e} e^{\frac{2}{p} h(\bm{X} \vert \bm{Z}=\bm{z})}\right] \geq \frac{p}{2 \pi e} e^{\frac{2}{p} h(\bm{X} \vert \bm{Z})}.
\end{equation}
(\ref{proof_lemma_5}) holds with equality if and only if $h(\bm{X} \vert \bm{Z}=\bm{z})$ is constant. Based on (\ref{proof_lemma_1}), (\ref{proof_lemma_4}) and (\ref{proof_lemma_5}), we finally obtain
$$
    \mathbb{E}\left[\left\lVert \bm{X} - \hat{\bm{X}}\right\rVert ^2\right] \geq \frac{p}{2\pi e} e^{\frac{2}{p} h(\bm{X} \vert \bm{Z})}.
$$

From Lemma 1, the following corollary can be derived.

\textbf{Corollary 1.} \emph{For any continuous random vector $\bm{X}$, given observation $\bm{Z}$, side information $\bm{Y}$ and the corresponding estimation result $\hat{\bm{X}}^{\ast}$,}
\begin{equation}
    \label{corollary_1}
    \mathbb{E}\left[\left\lVert \bm{X} - \hat{\bm{X}^{\ast}}\right\rVert ^2\right] \geq \frac{p}{2\pi e} e^{\frac{2}{p} h(\bm{X} \vert \bm{Y}, \bm{Z})}.
\end{equation}
\emph{The inequality holds with equality if and only if the estimator is unbiased, $\bm{X} \vert \bm{Y}, \bm{Z}$ is Gaussian, and $\bm{\varSigma}\left(\bm{X} \vert \bm{Y}=\bm{y}, \bm{Z}=\bm{z}\right) = \sigma^2 \bm{I}_p$, where $\sigma$ is a constant independent of $\bm{y}$ and $\bm{z}$.}

To prove Corollary 1, it suffices to replace $\bm{X} \vert \bm{Z}$ in Lemma 1 with $\bm{X} \vert \bm{Y}, \bm{Z}$, and thus the proof will not be elaborated here.

According to Lemma 1 and Corollary 1, we can obtain the following theorem.

\textbf{Theorem 1.} \emph{For any continuous random vector $\bm{X}$, given observation $\bm{Z}$, side information $\bm{Y}$, the estimate $\hat{\bm{X}}$ obtained solely from $\bm{Z}$, and the estimate $\hat{\bm{X}}^\ast$ obtained when both $\bm{Y}$ and $\bm{Z}$ are known, the following holds:}
\begin{equation}
    \label{thm_1_1}
    \varXi\left(\hat{\bm{X}}^\ast\right) = \varXi\left(\hat{\bm{X}}\right) e^{-\frac{2}{p} I\left(\bm{X};\bm{Y}\vert\bm{Z}\right)},
\end{equation}
\emph{where $I\left(\bm{X};\bm{Y}\vert\bm{Z}\right) \geq 0$ represents the conditional mutual information, and $\varXi (\cdot)$ denotes the lower bound of the MSE given by Lemma 1 and Corollary 1, i.e., }
\begin{equation}
    \label{thm_1_2}
    \varXi\left(\hat{\bm{X}}\right)=\frac{p}{2\pi e} e^{\frac{2}{p} h(\bm{X} \vert \bm{Z})},
\end{equation}
\begin{equation}
    \label{thm_1_3}
    \varXi\left(\hat{\bm{X}}^\ast\right)=\frac{p}{2\pi e} e^{\frac{2}{p} h(\bm{X} \vert \bm{Y}, \bm{Z})}.
\end{equation}

\emph{Proof.}
\begin{equation}
    \label{proof_1}
    \begin{aligned}
        h(\bm{X} \vert \bm{Y}, \bm{Z}) = & h(\bm{X}) - I(\bm{X};\bm{Y}, \bm{Z}), \\
        = & h(\bm{X}) - I(\bm{X};\bm{Z}) - I(\bm{X};\bm{Y} \vert \bm{Z}), \\
        = & h(\bm{X} \vert \bm{Z}) - I(\bm{X};\bm{Y} \vert \bm{Z}). \\
    \end{aligned}
\end{equation}
Then we have
\begin{equation}
    \label{proof_2}
    \begin{aligned}
        \frac{p}{2\pi e} e^{\frac{2}{p} h(\bm{X} \vert \bm{Y}, \bm{Z})} = & \frac{p}{2\pi e} e^{\frac{2}{p} \left(h(\bm{X} \vert \bm{Z}) - I(\bm{X};\bm{Y} \vert \bm{Z})\right)}, \\
        = & \frac{p}{2\pi e} e^{\frac{2}{p} h(\bm{X} \vert \bm{Z})} e^{-\frac{2}{p} I(\bm{X};\bm{Y} \vert \bm{Z})}. \\
    \end{aligned}
\end{equation}
So
$$
    \varXi\left(\hat{\bm{X}}^\ast\right) = \varXi\left(\hat{\bm{X}}\right) e^{-\frac{2}{p} I\left(\bm{X};\bm{Y}\vert\bm{Z}\right)}.
$$

Theorem 1 shows that, when side information is applied, the lower bound of the MSE is reduced by a factor of $e^{-\frac{2}{p} I\left(\bm{X};\bm{Y}\vert\bm{Z}\right)}$. $I\left(\bm{X};\bm{Y}\vert\bm{Z}\right)$ indicates that the lower bound of the MSE decreases when the side information $\bm{Y}$ provides information about $\bm{X}$ beyond the observation $\bm{Z}$; otherwise, the lower bound remains unchanged.

Although Theorem 1 describes the change in the lower bound of the MSE, it still provides some guidance when the estimator is an MMSE estimator; in particular, when the MMSE decreases, we can analyze the source of the gain according to Theorem 1. For traditional estimators where the entire estimation process can be explicitly expressed, Theorem 1 indeed lacks value. However, for many current estimation problems that rely on deep learning, this theorem demonstrates significant value. As discussed before, for problems estimated using deep learning, when the loss function is set to MSE, the deep neural network can approximate the MMSE estimator. Therefore, when deep learning methods achieve significantly higher accuracy than traditional theoretical limits, we can use Theorem 1 to analyze possible sources of gain. 

In the following two subsections, we examine two specific forms of side information that commonly arise in deep learning-based Wi-Fi sensing---prior information about the target structure and temporal correlation across sequential CSI measurements---and show how each reduces the lower bound of the MSE within the framework of Theorem 1.

\subsection{Performance Gain Brought by Prior Information}

Prior information about the target structure can be interpreted as a form of side information $\bm{Y}$ in Theorem 1. In the context of HPE, this prior knowledge corresponds to the structural characteristics of the human body, such as limb lengths and joint connectivity. The model acquires this knowledge during training by learning from the data, effectively increasing $I(\bm{X};\bm{Y}\vert\bm{Z})$ and thereby reducing the MSE lower bound. We use indoor HPE as a concrete example to illustrate how this gain manifests in practice.

\begin{figure}[!ht]
    \centering                                    
    \includegraphics[width=3.2in]{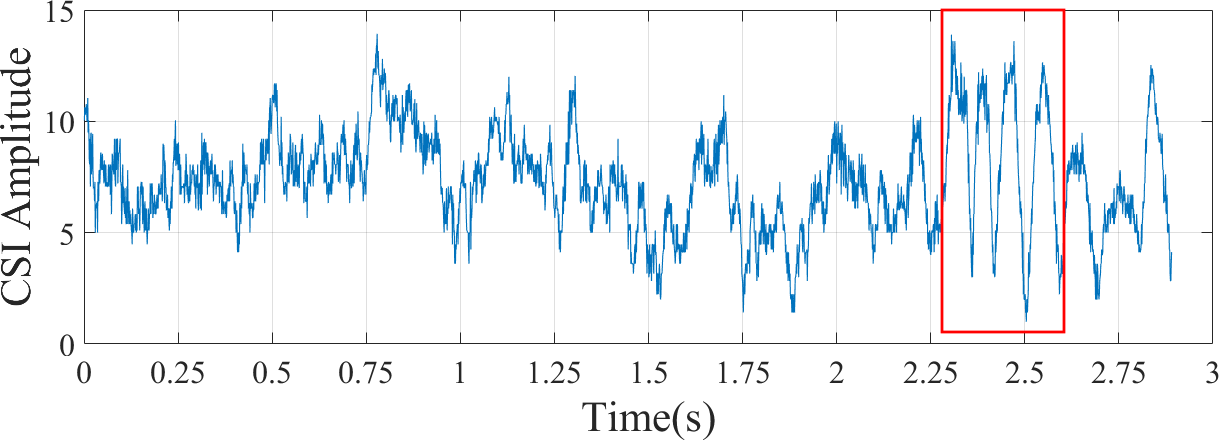}
    \caption{An example of the collected CSI. The area within the red box exhibits clear sinusoidal characteristics.}
    \label{CSI_sample} 
\end{figure}

The CSI-Speed \cite{wang2015understanding} model reveals the theoretical basis for using CSI signals for motion pattern recognition. Treating CSI as the sum of paths affected only by static objects and paths influenced by dynamic objects, we have
\begin{equation}
    \label{CSI_ori}
    h(f, t) = e^{-j\phi(f, t)} \left( h_{\text{s}}(f) + \sum_{l = 1}^{L_{\text{d}}} \alpha_l(f, t) e^{-j 2 \pi \frac{d_l(t)f}{c}}\right),
\end{equation}
where \( f \) denotes the carrier frequency, \( t \) represents time, $\phi(f, t)$ is the phase distortion caused by thermal noise and clock asynchronism at the receiver and transmitter, \( h_{\text{s}}(f) \) denotes the CSI of paths affected only by static objects, \( L_{\text{d}} \) is the number of paths influenced by moving objects, \( \alpha_l(f, t) \) represents the attenuation of the \( l \)-th path, \( d_l(t) \) denotes the length of the \( l \)-th path, and \( c \) is the speed of light. Then we have
\begin{equation}
    \label{CSI_spd}
    \begin{aligned}
        \| h(f, t) \|^2 =&\  \| h_{\text{s}}(f) \|^2 \\ 
        &+ 2 \sum_{l = 1}^{L_{\text{d}}} \| h_{\text{s}}(f) \alpha_l(f, t) \| \\
        &\quad \times \cos \left[ 2 \pi \left( \int_{0}^{t} \frac{v_l (\tau)}{c} f \, \mathrm{d}\tau 
           + \frac{d_l (0)f}{c} \right) + \varphi_{sl} \right] \\
        &+ \sum_{k = 1}^{L_{\text{d}}} \sum_{l = 1}^{L_{\text{d}}} \| \alpha_k(f, t) \alpha_l(f, t) \| \\
        &\quad \times \cos \left[ 2 \pi \frac{ ( d_k(t) - d_l(t) ) f }{c} + \varphi_{kl} \right],
    \end{aligned}
\end{equation}
where \( v_l(\tau) \) denotes the rate of change in the length of path \( l \), and \( \varphi_{sl} \), \( \varphi_{kl} \) represent constant phase shifts introduced by reflection. In indoor environments, due to strong reflections from walls, \( \| h_{\text{s}}(f)\| \gg \|\alpha_l(f, t) \| \); thus the third term in (\ref{CSI_spd}) can be neglected. Therefore, the CSI in indoor environments can be modeled as a superposition of a static component and multiple dynamic sinusoidal components. This sinusoidal behavior is clearly observed in the actually collected CSI data over time, as demonstrated in Fig. \ref{CSI_sample}.

Consequently, different velocity combinations manifest in the CSI as distinct superpositions of sinusoidal signals. The characteristics of these superimposed sinusoidal signals can be utilized to recognize the target's actions, forming the theoretical basis for deep learning-based action classification tasks. 

In regression tasks, these features can still yield performance gains. For HPE, deep learning does not need to determine the precise coordinates of each joint. Instead, if motion pattern recognition can be achieved, the model can generate a human pose corresponding to the action category based on prior information about human limb proportions learned during training. We refer to this gain as the performance gain brought by prior information. A similar form of gain is also observed in related fields, such as deep learning-based super-resolution in image processing. The experiments in Section IV-D present evidence supporting this claim.

\subsection{Performance Gain Brought by Temporal Correlation}

Temporal correlation constitutes another specific form of side information within the framework of Theorem 1. Here, past CSI measurements serve as the side information $\bm{Y}$, i.e., $\bm{Y}=\bm{Z}_{t-1\cdots 0}$, where $\bm{Z}_{t-1\cdots 0}$ means $\bm{Z}_{t-1}, \bm{Z}_{t-2}, \cdots, \bm{Z}_{0}$, and $\bm{Z}_{t}$ denotes $\bm{Z}$ at time $t$. Then the following corollary can be derived.

\textbf{Corollary 2.} Assume that the pair of random processes $\left\{\bm{X}_t, \bm{Z}_t\right\}_{t \geq 0}$ satisfies a hidden Markov model (HMM), i.e.,
\begin{itemize}
    \item $\bm{X}_{t-1} \to \bm{X}_{t}$, 
    \item $\bm{X}_{t-1} \to \bm{Z}_{t-1}$, $\bm{X}_{t} \to \bm{Z}_{t}$, 
\end{itemize}
where $\bm{X}_t$ denotes the state to be estimated at time $t$, $\bm{Z}_t$ denotes the observation at time $t$. Then, given the historical observations, the lower bound of the MSE is strictly reduced, if and only if there exists a pair $(x_t, z_t) \in \mathrm{supp}\,P_{\bm{X}_t, \bm{Z}_t}$ such that
\begin{equation}
    \label{corollary_2_3}
    P\left(\bm{Z}_{t-1 \cdots 0} \vert \bm{X}_t = x_t\right) \neq P\left(\bm{Z}_{t-1 \cdots 0} \vert \bm{Z}_t = z_t\right), 
\end{equation}
where $\mathrm{supp}\,P_{\bm{X}_t, \bm{Z}_t}$ denotes the support of the joint distribution of $\bm{X}_t$ and $\bm{Z}_t$.

\emph{Proof.} We first prove the forward direction, i.e., that condition (\ref{corollary_2_3}) entails the strict reduction of the lower bound. According to Theorem 1, we have
\begin{equation}
    \label{corollary_2_1}
    \varXi\left(\hat{\bm{X}}^\ast_t\right) = \varXi\left(\hat{\bm{X}}_t\right) e^{-\frac{2}{p} I\left(\bm{X}_t;\bm{Z}_{t-1 \cdots 0}\vert\bm{Z}_t\right)}.
\end{equation}
By the observation independence of the hidden Markov model, the observations are conditionally independent given the states, i.e.,
\begin{equation}
    \label{corollary_2_4}
    P(\bm{Z}_{t \cdots 0} \vert \bm{X}_{t \cdots 0}) = \prod_{k=0}^{t} P(\bm{Z}_k \vert \bm{X}_k).
\end{equation}
Marginalizing $\bm{X}_{t-1 \cdots 0}$, we obtain
\begin{equation}
    \label{corollary_2_5}
    \begin{aligned}
        P&(\bm{Z}_{t \cdots 0} \vert \bm{X}_t) \\
        &= \sum_{\bm{x}_{t-1 \cdots 0}} P(\bm{Z}_{t \cdots 0} \vert \bm{x}_{t-1 \cdots 0}, \bm{X}_t) P(\bm{x}_{t-1 \cdots 0} \vert \bm{X}_t) \\
        &= P(\bm{Z}_t \vert \bm{X}_t) \sum_{\bm{x}_{t-1 \cdots 0}} P(\bm{Z}_{t-1 \cdots 0} \vert \bm{x}_{t-1 \cdots 0}) P(\bm{x}_{t-1 \cdots 0} \vert \bm{X}_t) \\
        &= P(\bm{Z}_t \vert \bm{X}_t)P(\bm{Z}_{t-1 \cdots 0} \vert \bm{X}_t),
    \end{aligned}
\end{equation}
i.e., 
\begin{equation}
    \label{corollary_2_6}
    \bm{Z}_{t-1 \cdots 0} \perp \bm{Z}_t \mid \bm{X}_t,
\end{equation}
where $\perp$ denotes statistical independence. Equivalently,
\begin{equation}
    \label{corollary_2_7}
    P(\bm{Z}_{t-1 \cdots 0} \vert \bm{X}_t, \bm{Z}_t) = P(\bm{Z}_{t-1 \cdots 0} \vert \bm{X}_t).
\end{equation}
Suppose that $I\left(\bm{X}_t;\bm{Z}_{t-1 \cdots 0}\vert\bm{Z}_t\right) = 0$, which is equivalent to $\bm{X}_t \perp \bm{Z}_{t-1 \cdots 0} \mid \bm{Z}_t$, i.e., 
\begin{equation}
    \label{corollary_2_8}
    P(\bm{Z}_{t-1 \cdots 0} \vert \bm{X}_t, \bm{Z}_t) = P(\bm{Z}_{t-1 \cdots 0} \vert \bm{Z}_t). 
\end{equation}
Combining this with (\ref{corollary_2_7}), we obtain
\begin{equation}
    \label{corollary_2_9}
    P(\bm{Z}_{t-1 \cdots 0} \vert \bm{X}_t = x_t) = P(\bm{Z}_{t-1 \cdots 0} \vert \bm{Z}_t = z_t)
\end{equation}
for every pair $(x_t, z_t) \in \mathrm{supp}\,P_{\bm{X}_t, \bm{Z}_t}$, which contradicts condition (\ref{corollary_2_3}). Hence $I\left(\bm{X}_t;\bm{Z}_{t-1 \cdots 0}\vert\bm{Z}_t\right) > 0$, and the strict inequality
\begin{equation}
    \label{corollary_2_10}
    \varXi\left(\hat{\bm{X}}^\ast_t\right) < \varXi\left(\hat{\bm{X}}_t\right)
\end{equation}
follows from (\ref{corollary_2_1}). Since every step of the above derivation is reversible, the converse also holds.

A typical example for Corollary 2 is that, for Wi-Fi CSI-based HPE problems, it is often difficult for the model to distinguish the subject's orientation (e.g., facing toward or away from the receiver), but if the subject has previously walked, the orientation can be determined based on the walking direction.

We now discuss the range of applicability of Corollary 2. The proof above establishes that (\ref{corollary_2_3}) is equivalent to \(I\left(\bm{X}_t;\bm{Z}_{t-1 \cdots 0}\vert\bm{Z}_t\right) > 0\). Therefore, (\ref{corollary_2_3}) essentially requires that past observations contain information about the current state that the current observation cannot provide.

This condition mainly covers two scenarios. First, when the distribution is faithful to the HMM graph structure, in which all conditional independencies follow from d-separation, the randomness of observations and state transitions ensures that (\ref{corollary_2_3}) holds. Second, even though faithfulness fails, when the state transition is deterministic, (\ref{corollary_2_3}) still holds if past observations can distinguish different current states. This includes injective transition mappings where observations distinguish different states, and non-injective mappings where observations distinguish different successor states. These two scenarios cover almost all practically meaningful situations.

The strict reduction fails if and only if $P(\bm{Z}_{t-1 \cdots 0}\vert\bm{X}_t=x)=P(\bm{Z}_{t-1 \cdots 0}\vert\bm{Z}_t=z)$ holds for every $(x,z) \in \mathrm{supp}\,P_{\bm{X}_t,\bm{Z}_t}$. This occurs, for instance, when the current state is nearly independent of its past, so that the past observations carry no information about it; when the observation cannot distinguish the states; or when the observation is noiseless and injective, so that the current observation determines the state exactly and the entire history is redundant. In any case, the bound in (\ref{corollary_2_1}) never increases, since the conditional mutual information is always non-negative.

\section{Deep Learning-Based Real-Time Wi-Fi Sensing and Visualization System}

Building on our theoretical insights, we develop a deep learning-based real-time Wi-Fi sensing and visualization system. In this section, we provide a detailed description of the system’s design.

\begin{figure*}[!ht]
	\centering
	\includegraphics[width=6.9in]{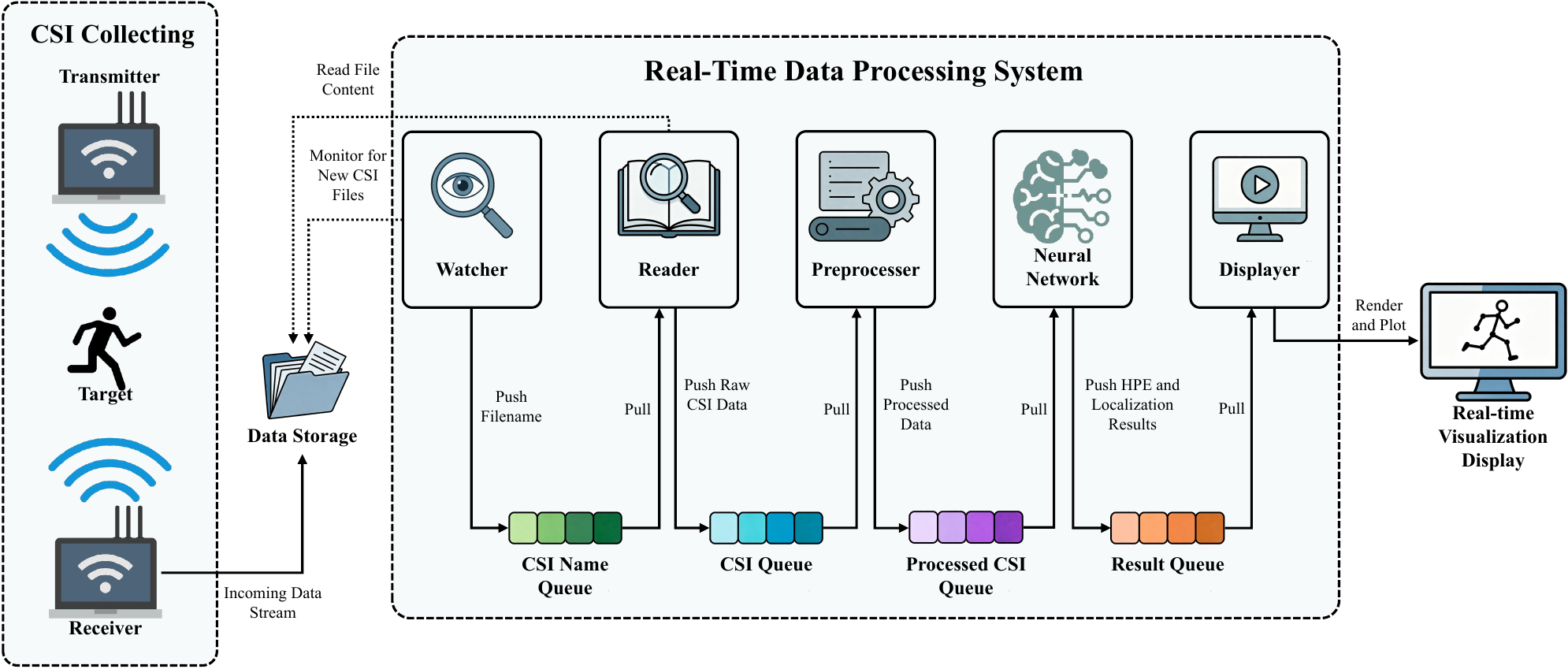}
	\caption{Schematic diagram of the deep learning-based real-time Wi-Fi sensing and visualization system.}
	\label{system}
\end{figure*}

\subsection{System Overview}

The proposed deep learning-based real-time Wi-Fi sensing and visualization system is illustrated in Fig. \ref{system}, which estimates human pose and location from Channel State Information (CSI) extracted from Wi-Fi signals. CSI data are collected in an indoor environment using the Linux 802.11n CSI Tool~\cite{halperin2011tool} and stored. The real-time processing system monitors local file updates, reads the latest CSI files, preprocesses the data in real time, and feeds them into a neural network for human pose estimation and localization. The results are plotted and displayed in real time. Note that the location of a person is defined as the center of the torso, while the pose is represented by the coordinates of each joint relative to this location.

\subsection{CSI Data}

CSI characterizes how signals propagate from the transmitter to the receiver through multiple paths \cite{wang2015understanding}. If a signal is transmitted through $L_{(n, m)}$ different propagation paths, the CSI characterizing the channel can be represented as
\begin{equation}
    \label{CSI_ft}
    h_{(n, m)}(f, t) = e^{-j\phi(f, t)} \sum_{l = 1}^{L_{(n, m)}} \alpha_{l(n, m)}(f, t) e^{-j2 \pi f \tau_{l(n, m)}(t)},
\end{equation}
where the subscript \((n, m)\) denotes the \(n\)-th receive antenna and the \(m\)-th transmit antenna, and $\tau_{l(n, m)}(t)$ is the propagation delay. If there are $N_\text{t}$ transmitting antennas and $N_\text{r}$ receiving antennas, the CSI can be represented as 
\begin{equation}
    \label{CSI_mat}
    \bm{h}(f, t) = 
    \begin{bmatrix}
        h_{(1, 1)}(f, t)          & \cdots & h_{(1, N_\text{t})}(f, t)          \\
        \vdots                    & \ddots & \vdots                             \\
        h_{(N_\text{r}, 1)}(f, t) & \cdots & h_{(N_\text{r}, N_\text{t})}(f, t) \\
    \end{bmatrix}.
\end{equation}

Notably, the influence of $\phi$ is highly stochastic, and in indoor environments, reflections from the target are often overwhelmed by those from walls. Both factors can significantly degrade the model performance. Therefore, preprocessing of the CSI data is essential to mitigate these adverse effects.

\subsection{CSI Preprocessing}
The preprocessing method employed in this work follows the framework introduced in \cite{qian2018widar2}, leveraging the CSI from a single receiving antenna as a reference to reconstruct the dynamic components present in the CSI from all other antennas. However, since the aforementioned study utilized systems with a single transmitting antenna and multiple receiving antennas, our subsequent derivation in this section demonstrates that directly applying this method to a more general system with $N_\text{t}$ transmitting antennas would result in discarding $N_\text{t}-1$ elements of the CSI matrix. To address this limitation, we have introduced necessary modifications to the method. 

We begin by reshaping the original CSI matrix $\bm{h}(f, t)$ into an $N_\text{t} N_\text{r} \times 1$ column vector $\bm{h}'(f, t)$. We have
\begin{equation}
	\label{CSI_vect}
	\begin{aligned}
		\bm{h}'(f, t) = & \left[ h_{(1, 1)}(f, t) \cdots h_{(1, N_{\text{t}})}(f, t), \right.\\
		& \left. h_{(2, 1)}(f, t) \cdots h_{(2, N_{\text{t}})}(f, t) \cdots h_{(N_\text{r}, N_\text{t})}(f, t) \right]^T, \\
	\end{aligned}
\end{equation}
Treating $\bm{h}'(f, t)$ as being composed of both $\bm{h}_{\text{s}}(f)$, which arises solely from paths affected by static objects, and $\bm{h}_{\text{d}}(f, t)$, which arises from paths influenced by moving objects, i.e.,
\begin{equation}
    \label{CSI}
    \bm{h}'(f, t) = (\bm{h}_{\text{s}}(f) + \bm{h}_{\text{d}}(f, t))e^{-j\phi(f, t)}.
\end{equation}
Then we have 
\begin{equation}
    \label{conj_mul}
    \begin{split}
        \bm{h}'(f, t) \bm{h}'^\dagger (f, t) = & \bm{h}_{\text{s}}(f) \bm{h}_{\text{s}}^\dagger (f) + \bm{h}_{\text{s}}(f) \bm{h}_{\text{d}}^\dagger (f, t) \\
                                             & + \bm{h}_{\text{d}}(f, t) \bm{h}_{\text{s}}^\dagger (f) + \bm{h}_{\text{d}}(f, t) \bm{h}_{\text{d}}^\dagger (f, t),
    \end{split}
\end{equation}
where $\dagger$ means conjugate transpose. Let the element of matrix $\bm{A}$ in row n and column m be $\bm{A}_{(n, m)}$, and the $n\text{th}$ element of vector $\bm{B}$ be $\bm{B}_{(n)}$. In indoor environments, the reflections from walls are very strong, so for all $0 < n \leq N_{\text{t}} N_{\text{r}}$, $\left\lVert \bm{h}_{\text{s}(n)} \right\rVert \gg \left\lVert \bm{h}_{\text{d}(n)} \right\rVert$. Then we have 
\begin{equation}
    \label{conj_mul_2}
    \begin{split}
        & \left\lvert(\bm{h}_{\text{s}}(f) \bm{h}_{\text{s}}^\dagger (f) + \bm{h}_{\text{s}}(f) \bm{h}_{\text{d}}^\dagger (f, t) + \bm{h}_{\text{d}}(f, t) \bm{h}_{\text{s}}^\dagger (f))_{(n, m)}\right\rvert \\
        & \gg \left\lvert(\bm{h}_{\text{d}}(f, t) \bm{h}_{\text{d}}^\dagger (f, t))_{(n, m)}\right\rvert , 
    \end{split}
\end{equation}
where $0 < n \leq N_{\text{r}}, 0 < m \leq N_{\text{t}}$. So
\begin{equation}
    \label{conj_mul_3}
    \bm{h}'(f, t) \bm{h}'^\dagger (f, t) \approx \bm{h}_{\text{s}}(f) \bm{h}_{\text{s}}^\dagger (f) + \bm{h}_{\text{s}}(f) \bm{h}_{\text{d}}^\dagger (f, t) + \bm{h}_{\text{d}}(f, t) \bm{h}_{\text{s}}^\dagger (f),
\end{equation}
where $\bm{h}_{\text{s}}(f) \bm{h}_{\text{s}}^\dagger (f)$ can be eliminated by a high-pass filter (HPF), i.e.,
\begin{equation}
    \label{conj_mul_4}
    \begin{split}
        \bm{D}(f, t) & = \text{HPF}(\bm{h}'(f, t) \bm{h}'^\dagger (f, t)) \\
                     & \approx \bm{h}_{\text{s}}(f) \bm{h}_{\text{d}}^\dagger (f, t) + \bm{h}_{\text{d}}(f, t) \bm{h}_{\text{s}}^\dagger (f).
    \end{split}
\end{equation}
For $\bm{h}_{\text{d}}(f, t)$, which we care about, we cannot obtain it independently from (\ref{conj_mul_4}).

Take $N_{\text{ref}}$ elements in $\bm{h}'(f, t)$ for reference, where $1 \leq N_{\text{ref}} < N_{\text{t}} N_{\text{r}}$. For convenience, we assume that the reference elements are the first $N_{\text{ref}}$ elements of $\bm{h}'(f, t)$. In practice, the order of the elements of $\bm{h}'(f, t)$ can be adjusted so that the selected reference elements always appear at the beginning. To highlight the dynamic component of $\bm{h}'(f, t)$, we introduce a vector $\bm{\beta}$ of dimension $N_{\text{ref}} \times 1$, then
\begin{equation}
    \label{CSI_mod}
        \bm{h}'(f, t) + \binom{\bm{\beta}}{\bm{0}} = \left[(\bm{h}_{\text{s}}(f) + \binom{\bm{\beta}}{\bm{0}}e^{j\phi(f, t)}) + \bm{h}_{\text{d}}(f, t)\right] e^{-j\phi(f, t)}.
\end{equation}
In the following derivation, we will see that when $\bm{\beta}$ is chosen appropriately, the dynamic components of all elements in $\bm{h}'(f, t)$ can be highlighted except for the reference elements. We have
\begin{equation}
    \label{conj_mul_mod_2}
    \begin{split}
        \bm{D}'(f, t) = & \text{HPF}\left[\left(\bm{h}'(f, t) + \binom{\bm{\beta}}{\bm{0}}\right) \left(\bm{h}'(f, t) + \binom{\bm{\beta}}{\bm{0}}\right)^\dagger\right] \\
                      = & \left(\bm{h}_{\text{s}}(f) + \binom{\bm{\beta}}{\bm{0}}e^{j\phi(f, t)}\right) \bm{h}_{\text{d}}^\dagger (f, t) \\
                        & + \bm{h}_{\text{d}}(f, t) \left(\bm{h}_{\text{s}}(f) + \binom{\bm{\beta}}{\bm{0}}e^{j\phi(f, t)}\right)^\dagger \\
                        & + \bm{h}_{\text{d}}(f, t) \bm{h}_{\text{d}}^\dagger (f, t).
    \end{split}
\end{equation}

\begin{figure*}[!ht]
	\centering
	\includegraphics[width=6.4in]{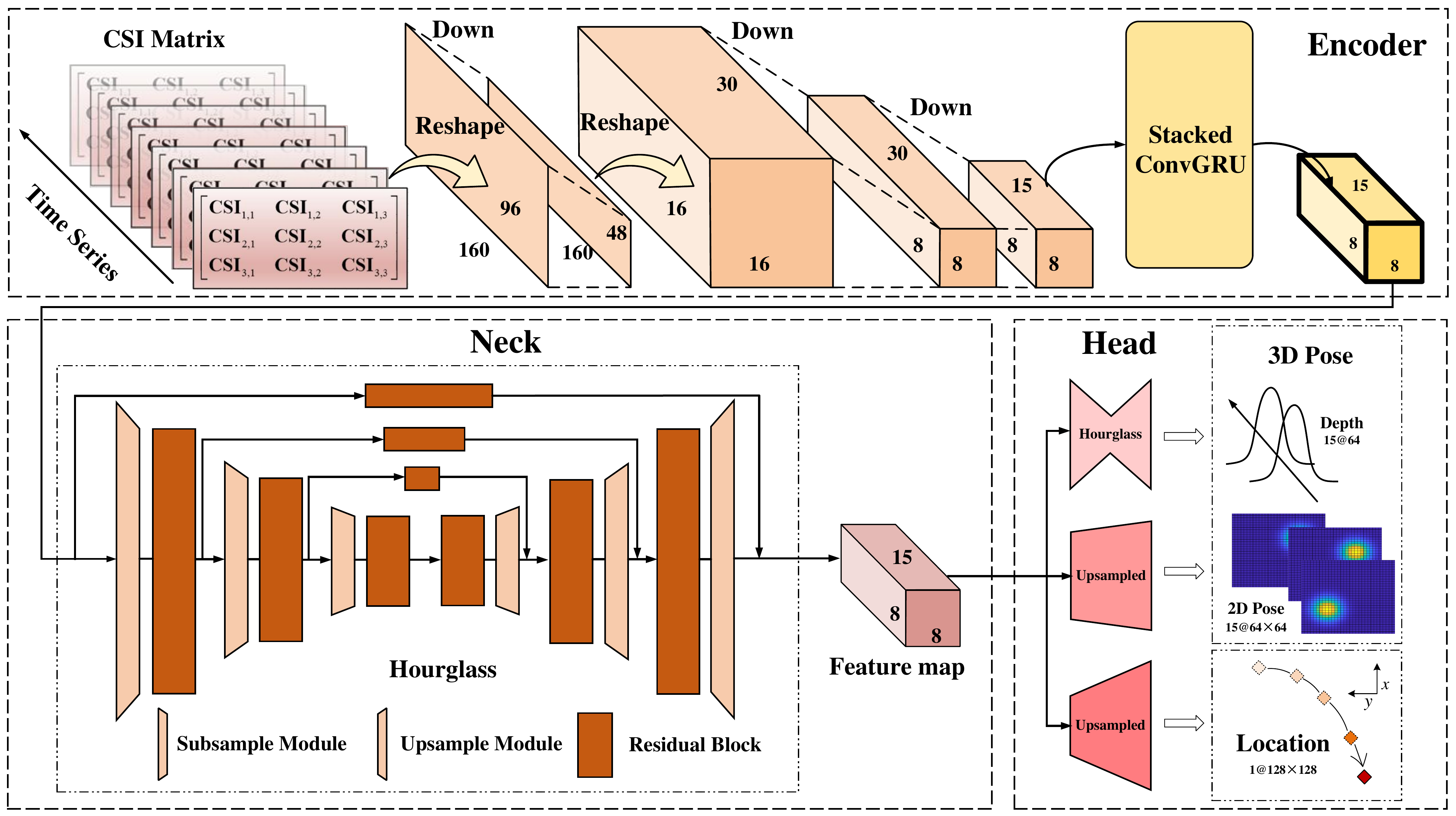}
	\caption{Illustration of the network structure, where we incorporate a hyperparameter in the stacked ConvGRU layers to control the maximum processable sequence length. The depth head and 2D pose estimation head work jointly to achieve 3D human pose estimation.}
	\label{Network}
\end{figure*}

Assume $\left\lvert\bm{\beta}_{(m)}\right\rvert \gg \left\lvert\bm{h}_{\text{s}(n)}(f)\right\rvert $; then we have
\begin{equation}
    \label{conj_mul_mod_3}
    \bm{D}'_{(n, m)}(f, t) \approx (\bm{h}_{\text{s}(m)}(f) + \bm{\beta}_{(m)} e^{j\phi(f, t)})^\ast \bm{h}_{\text{d}(n)}(f, t),
\end{equation}
where $ 1 \leq m \leq N_{\text{ref}} < n \leq N_{\text{t}} N_{\text{r}}$, and $\ast$ means conjugate. Let $\bm{\beta}_{(m)}$ have the same phase as $\bm{h}'_{(m)}(f, t)$, $1 \leq m \leq N_{\text{ref}}$. Since $\left\lVert \bm{h}_{\text{s}(l)} \right\rVert  \gg \left\lVert \bm{h}_{\text{d}(l)} \right\rVert$, $0 < l \leq N_{\text{t}} N_{\text{r}}$, then $\frac{\bm{\beta}_{(m)} e^{j\phi(f, t)}}{\left\lVert \bm{\beta}_{(m)} e^{j\phi(f, t)} \right\rVert } \approx \frac{\bm{h}_{\text{s}(m)}(f)}{\left\lVert \bm{h}_{\text{s}(m)}(f) \right\rVert }$. In this case, the effect of $\bm{\beta}_{(m)}$ on $\bm{h}_{\text{s}(m)}(f)$ can be considered as increasing the modulus of $\bm{h}_{\text{s}(m)}(f)$, and the phase change of $\bm{h}_{\text{s}(m)}(f)$ is small. Let $\bm{h}'_{\text{s}(m)}(f) = \bm{h}_{\text{s}(m)}(f) + \bm{\beta}_{(m)} e^{j\phi(f, t)}$; then, if $ 1 \leq m \leq N_{\text{ref}} < n \leq N_{\text{t}} N_{\text{r}}$, 
\begin{equation}
    \label{conj_mul_mod_4}
    \bm{D}'_{(n, m)}(f, t) \approx \bm{h}'^{\, \ast}_{\text{s}(m)}(f) \bm{h}_{\text{d}(n)}(f, t),
\end{equation}
According to (\ref{conj_mul_mod_4}), when $1 \leq m \leq N_{\text{ref}} < n \leq N_{\text{t}} N_{\text{r}}$, $\bm{D}'_{(n, m)}(f, t)$ is linear in the dynamic component. In this case, in the original $N_\text{t} N_\text{r} \times 1$ CSI vector $\bm{h}'(f, t)$, we obtain $N_{\text{t}} N_{\text{r}} - N_{\text{ref}}$ elements of $\bm{D}'_{(n, m)}(f, t)$ that are approximately linear in the dynamic component at the cost of abandoning $N_{\text{ref}}$ reference elements. Since the dynamic component is critical for HPE, we aim to maximize the number of elements of $\bm{D}'_{(n, m)}(f, t)$ that are linear in the dynamic component. To achieve this, we take $N_{\text{ref}} = 1$ and set the reference element as the element with the smallest average magnitude in $\bm{h}'(f, t)$.

\subsection{Network Design}

The architecture of the proposed network (Fig. \ref{Network}) consists of an encoder for feature extraction and data compression, followed by a neck for intermediate processing, and a head for final pose estimation.

We employ an encoder to compress the input data and perform feature extraction. We begin by training an autoencoder, and the output of its encoder portion is then used as the input to the subsequent components of the network. In this encoder, we combine Convolutional Neural Networks (CNN) and MaxPooling layers to perform feature extraction and compression on the input CSI data. To enhance the encoder's sensitivity to temporal dependencies in the input data, we also incorporate two layers of Convolutional Gate Recurrent Units (ConvGRU). Furthermore, to investigate the impact of temporal correlation, the implemented ConvGRU architecture includes a hyperparameter that controls the maximum length of processable sequential data. When the number of continuously processed data samples is equal to the maximum processable sequence length, the model resets the hidden state of the ConvGRU to zero in order to clear the memory.

We employ an Hourglass network as the neck to further extract complex features embedded in the encoded CSI data. The network is equipped with three dedicated heads: one for 2D pose estimation, another for depth estimation, and the third for localization. The 2D pose estimation head and the depth estimation head jointly form the 3D HPE capability by integrating their respective outputs. To accelerate model convergence, we follow a common practice in computer vision by having each head output heatmaps instead of directly predicting the coordinates.

\subsection{Loss Function}

The loss function is composed of a weighted sum of the individual loss functions from each head, i.e.,
\begin{equation}
    \label{loss_total}
    \mathcal{L} = a \left[ b \mathcal{L}_{\text{Depth}} + \left(1-b\right) \mathcal{L}_{\text{HPE}} \right] + \left(1-a\right) \mathcal{L}_{\text{Location}}, 
\end{equation}
where $\mathcal{L}_{\text{head}}$ denotes the loss function of a head, with weights $a$ and $b$ satisfying $a, b \in \left(0, 1\right)$.

The loss function for each head comprises two components: a peak location error and a pixel-wise error. To combine the errors of these two quantities with different physical dimensions, we use the normalized mean square error (NMSE), which is a dimensionless measure. We denote the output of a head as $\widehat{\bm{m}}_{\text{head}}$ and its corresponding ground truth as $\bm{m}_{\text{head}}$. Thus, the error for each head can be expressed as
\begin{equation}
    \label{loss_depth}
    \mathcal{L}_{\text{Depth}} = \frac{\text{MSE}\left(\text{argmax}\left(\widehat{\bm{m}}_{\text{Depth}}\right)\right)}{\sigma^2\left(\text{argmax}\left(\bm{m}_{\text{Depth}}\right)\right)} c + \frac{\text{MSE}\left(\widehat{\bm{m}}_{\text{Depth}}\right)}{\sigma^2\left(\bm{m}_{\text{Depth}}\right)} \left(1-c\right), 
\end{equation}
\begin{equation}
    \label{loss_hpe}
    \mathcal{L}_{\text{HPE}} = \frac{\text{MSE}\left(\text{argmax}\left(\widehat{\bm{m}}_{\text{HPE}}\right)\right)}{\sigma^2\left(\text{argmax}\left(\bm{m}_{\text{HPE}}\right)\right)} c + \frac{\text{MSE}\left(\widehat{\bm{m}}_{\text{HPE}}\right)}{\sigma^2\left(\bm{m}_{\text{HPE}}\right)} \left(1-c\right), 
\end{equation}
\begin{equation}
    \label{loss_location}
    \begin{split}
        \mathcal{L}_{\text{Location}} = & \frac{\text{MSE}\left(\text{argmax}\left(\widehat{\bm{m}}_{\text{Location}}\right)\right)}{\sigma^2\left(\text{argmax}\left(\bm{m}_{\text{Location}}\right)\right)} c \\
                                        & + \frac{\text{MSE}\left(\widehat{\bm{m}}_{\text{Location}}\right)}{\sigma^2\left(\bm{m}_{\text{Location}}\right)} \left(1-c\right), 
    \end{split}
\end{equation}
where \(\mathrm{MSE}(\cdot)\) denotes the mean squared error, \(\sigma^2(\cdot)\) represents the variance, \(\mathrm{argmax}(\cdot)\) indicates the operation of finding the position of the maximum value for each channel, and \(c\) is a weighting coefficient satisfying \(c \in [0, 1]\), which can be adjusted based on the specific objectives at different stages of training. During the initial training stages, a lower value is assigned to \(c\) to facilitate rapid convergence. In later training, \(c\) is increased to refine the accuracy of peak localization.

\subsection{Real-Time Data Processing Design}

To meet real-time requirements, the following three issues must be addressed:
\begin{itemize}
\item \textbf{Receiver system limitation}. The Linux 802.11n CSI Tool in the Wi-Fi sensing system operates on Ubuntu 14.04, which lacks support for running neural networks.
\item \textbf{Inefficient reading of numerous small files}. The CSI acquisition is implemented using the Linux 802.11n CSI Tool, which is written in C, whereas neural networks typically rely on Python. Thus, CSI data cannot be passed directly via memory; it must first be written to disk and then read by Python. Although CSI data are collected and saved at a rate of 1000 fps, reading it from the hard disk (HDD or SSD) is far slower, thus severely limiting the subsequent data processing frequency.
\item \textbf{High overhead of scanning filenames}. Reading CSI with Python requires scanning the entire directory to obtain all filenames before loading the oldest unread file to maintain sequential order. This repeated directory scanning further slows down the process.
\end{itemize}
Since the neural network structure is lightweight enough, its computational overhead is not the bottleneck for real-time performance. To resolve the above issues, we built a real-time sensing and visualization system.

To address the limitations of the receiver system, the receiver streams the captured CSI data in real time to a host computer. The host computer is a commodity portable personal computer, an HP OMEN 8 Pro, equipped with an NVIDIA GeForce RTX 3060 graphics card. All subsequent steps — including data reading, processing, and result visualization — are integrated into a single process on the host computer. To eliminate the overhead of scanning file names, we set up an inotify-based watcher. Whenever a new file is created, inotify generates a file update event, from which the file name can be directly obtained without scanning the entire directory. To address the inefficiency of reading a large number of small files, we adopted a multi-process and multi-thread parallel reading and processing scheme and employed queues to manage data flow between different processes. Data is pushed to the tail of the queue upon arrival, and the system pulls (and removes) data from the head of the queue when needed. This push/pull mechanism ensures that the order of data processing remains consistent with the original sequence in which the receiver collects the CSI data. 

\section{Experimental Results}

\subsection{Experimental Setup}

\begin{figure*}[!ht]
	\centering
	\includegraphics[width=6.6in]{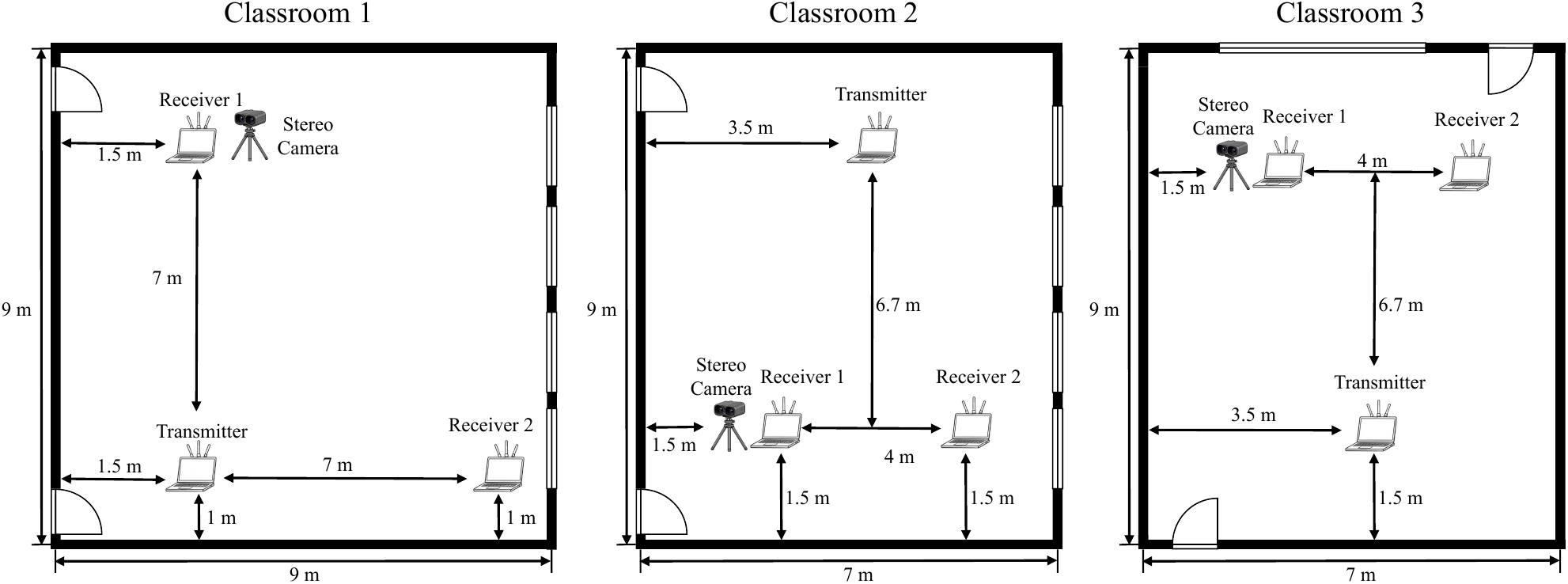}
	\caption{Illustration of the three classrooms used for data collection, showing the positions of the transmitter and receivers. During data collection, different propagation environments were created within the same classroom by varying the number and spatial distribution of scatterers.}
	\label{exp_env}
\end{figure*}

The proposed Wi-Fi sensing and visualization system operates on a single transceiver pair, but the data collection efficiency of one transceiver pair is relatively low. To improve efficiency, we used two receivers simultaneously during data collection. The transmitter and the two receivers were hosted on three personal computers, each equipped with an Intel 5300 NIC and configured with the Linux 802.11n CSI Tool \cite{halperin2011tool}. The two receivers were synchronized using the network time protocol, and the average synchronization error was below $10\ \text{ms}$. The transmitter and both receivers were each equipped with three antennas, and the receiving antennas faced the transmitting antennas. 

During data collection, the subjects performed the prescribed actions in the area between the transmitter and the receivers, while their 3D poses and CSI were recorded synchronously. The poses serve as ground truth and were captured by an Orbbec Gemini 336L structured-light stereo camera placed adjacent to one receiver and facing the same direction as its antennas; the resulting joint coordinates are consistent with their real-world positions.

The data collection system collects CSI over 30 subcarriers in the 5320 MHz band with a 10 MHz bandwidth at a rate of 1000 packets per second. Simultaneously, it captures video frames at a resolution of $1280 \times 800$ pixels and a frame rate of $30\ \text{fps}$, covering a field of view (FoV) of $90^\circ$ (horizontal) $\times$ $65^\circ$ (vertical) in front of the stereo camera.

As shown in Fig. \ref{exp_env}, CSI and 3D human pose data were collected in three different classrooms. To maximize the subjects' range of motion while ensuring that they remained within the FoV of the stereo camera, the configurations of the transmitter and receivers differed across the three classrooms. In addition, to make the CSI signals comparable across environments, we maintained the distance between the transmitter and each receiver at 7 m. Within the same classroom, different propagation environments were created by varying the quantity and spatial distribution of scatterers, yielding a total of 45 indoor propagation environments from which data were collected.

A total of six participants with varying body shapes, heights, and genders were involved in the data collection. Each participant performed four types of actions: walking, horizontally raising the left arm, horizontally raising the right arm, and horizontally raising both arms. The sample size for each action was balanced. 

\subsection{Preprocessing}

The dimension of the CSI obtained per packet is $3 \times 3 \times 30$. Following the approach described in Section III-C, we preprocessed the CSI and retained the reference signal, resulting in a signal of dimension \(9 \times 30\). Since this signal is complex-valued, both its amplitude and phase are separately fed into the network. Thus, the final dimension of the preprocessed CSI is \(2 \times 9 \times 30\) per packet.

Collected video frames are processed in two stages:
\begin{itemize}
    \item In the first stage, a 3D multi-person pose estimation (3DMPPE) method \cite{moon2019camera} is used to extract the 3D human skeletal structure. We select 16 major joints that represent the essential human pose information.
    \item In the second stage, the Semi-Global Block Matching (SGBM) algorithm \cite{SGBM} is applied to compute the depth map of the current frame. Then, the YOLOv10 model \cite{wang2024yolov10} detects the pixel coordinates of the human torso center in the frame. These coordinates are used to index the depth map, retrieving the world coordinates (i.e., coordinates relative to the camera) of the torso center.
\end{itemize}
Using this torso center location, the 3D human pose obtained in the first stage is adjusted to derive the joint coordinates in real physical space. The above video frame processing pipeline achieves centimeter-level average error between the captured joint positions and their ground-truth locations.

\begin{figure*}[!ht]
    \centering
    \subfloat[]{\includegraphics[width=3.1in]{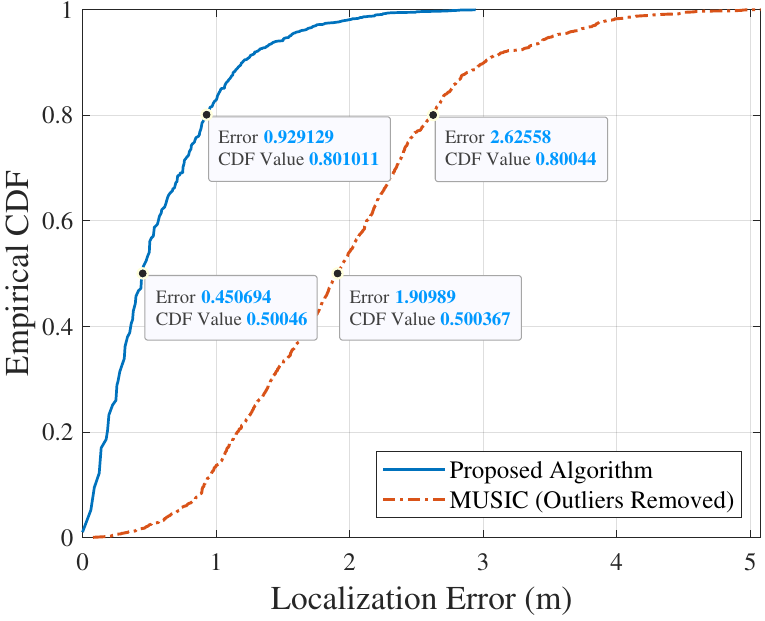}%
        \label{CDF_locate}}
    \hfil
    \subfloat[]{\includegraphics[width=3.1in]{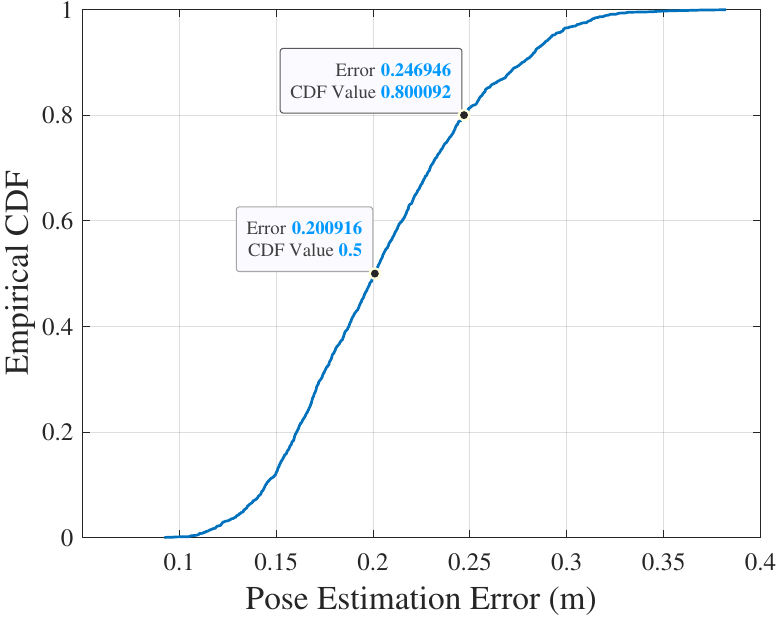}%
        \label{CDF_HPE}}
    \caption{(a) CDF of the localization error of the proposed system and of the MUSIC baseline under the same single-transceiver-pair setting. The MUSIC curve is plotted after excluding estimates that fall outside the room, which are spurious peaks caused by the antenna spacing of our testbed. (b) CDF of the human pose estimation error of the proposed system.}
    \label{CDF}
\end{figure*}

Since two receivers were employed while only one camera was available, the camera was co-located with one receiver. The captured pose data therefore require additional processing before it can serve as training targets for the CSI data from both receivers.

First, the ground plane is estimated to establish a consistent world coordinate system. For each data collection session, the lowest point of each pose frame is extracted, and a plane $ax + y + cz = b$ is fitted to these points via least squares, with normal vector $\bm{n} = (a, 1, c)^\mathrm{T}$. A rotation matrix $\bm{R}$ and translation vector $\bm{T}$ are constructed to align $\bm{n}$ with the $y$-axis and place the origin on the ground plane. After this step, all pose coordinates are expressed in a world-aligned frame independent of the camera's tilt and height.

Second, since the camera was attached to one receiver, the world-aligned pose represents the subject from its perspective. To obtain the corresponding ground truth for the other receiver's CSI data, the world-aligned coordinates $\bm{p}$ are transformed into the other receiver's local frame via a rigid transformation:
\begin{equation}
    \label{rigid_transform}
    \bm{p}' = \bm{R}_{\text{mod}}\,\bm{p} + \bm{t}_{\text{mod}},
\end{equation}
where $\bm{R}_{\text{mod}}$ is the rotation matrix determined by the relative antenna orientation between the two receivers, and $\bm{t}_{\text{mod}} \in \mathbb{R}^3$ is the spatial offset between their antenna array centers. This ensures that the CSI data from both receivers are paired with ground truth poses expressed in their respective local coordinate frames during training.

\subsection{Data Alignment and Segmentation}

The captured human pose data and the video frames share the same frame rate of $30 \text{ Hz}$, whereas the CSI is sampled at $1000 \text{ Hz}$. Temporal alignment between the pose data and CSI is achieved using timestamps. For each pose instance, we assign the $32$ temporally nearest CSI frames, thereby forming a data sample comprising a CSI tensor of dimension \(32 \times 2 \times 9 \times 30\) as input to the network, with the corresponding human pose serving as the ground truth.

A total of $24208$ data samples were collected. The dataset was partitioned into $18768$ samples for training, $3264$ for validation, and $2176$ for testing. To evaluate the influence of temporal correlation, the data were arranged chronologically and divided into contiguous blocks. In the collected dataset, the length of continuous data sequences ranges from a minimum of 68 to a maximum of 90. We set the block length to 17 continuous samples (i.e., one-fourth of 68), meaning that a block has dimensions $17 \times 32 \times 2 \times 9 \times 30$. After segmentation, data shorter than one block length are discarded. During both training and testing, a batch size of $16$ such blocks was utilized. Under such a design, data from different blocks in a batch are processed in parallel, while data within the same block are sequentially input into the network in chronological order to ensure that the ConvGRU maintains an effective memory of past inputs from the same block.

\subsection{Experimental Results}

\begin{table*}[!ht]
    \centering
    \caption{\label{MAE_joints}The mean absolute error (MAE) of HPE.}
    \begin{tabular}{|c|c|c|c|c|c|c|c|}
        \hline
        Joint & Thorax & Right Shoulder & Right Elbow & Right Wrist & Left Shoulder & Left Elbow & Left Wrist \\
        \hline
        MAE (m) & 0.0886 & 0.1565         & 0.3678      & 0.4763      & 0.1799        & 0.3517     & 0.4856     \\
        \hline
    \end{tabular}

    \begin{tabular}{|c|*{9}{c|}}
        \hline
        Joint & Right Hip & Right Knee & Right Ankle & Left Hip & Left Knee & Left Ankle & Pelvis & Head & MPJPE \\
        \hline
        MAE (m) & 0.1513 & 0.1599 & 0.2197 & 0.1335 & 0.1393 & 0.1787 & 0.0872 & 0.1069 & 0.2189 \\
        \hline
    \end{tabular}
\end{table*}

To demonstrate the advantage of deep learning over traditional algorithms, we compare our model with the classical radar super-resolution algorithm MUSIC on the indoor localization task. For the MUSIC baseline, the angle-of-departure (AoD) and the angle-of-arrival (AoA) are jointly estimated over all subcarriers, and the target position is obtained from the intersection of the corresponding AoD and AoA rays. The resulting cumulative distribution function (CDF) is shown in Fig. \ref{CDF}(a). Our algorithm achieves an average localization error of $0.6124\ \text{m}$, with approximately $50\%$ of the localization results having an error below $0.45\ \text{m}$ and over $80\%$ below $0.93\ \text{m}$. For the MUSIC algorithm, since the antenna spacing of our system is not an integer multiple of half a wavelength, spurious peaks appear in its pseudospectrum, and about $45\%$ of the localization results fall outside the range of the room where the data were actually collected. Since these results are physically unrealistic, we treat them as estimation failures. After removing these out-of-range results, the average localization error of MUSIC is $1.95\ \text{m}$, with approximately $50\%$ of the results having an error below $1.9\ \text{m}$ and over $80\%$ below $2.63\ \text{m}$. These results show that, under the experimental setup of this work, the proposed deep learning-based system outperforms the conventional MUSIC algorithm.

\begin{figure}[!t]
    \centering
	\includegraphics[width=3.1in]{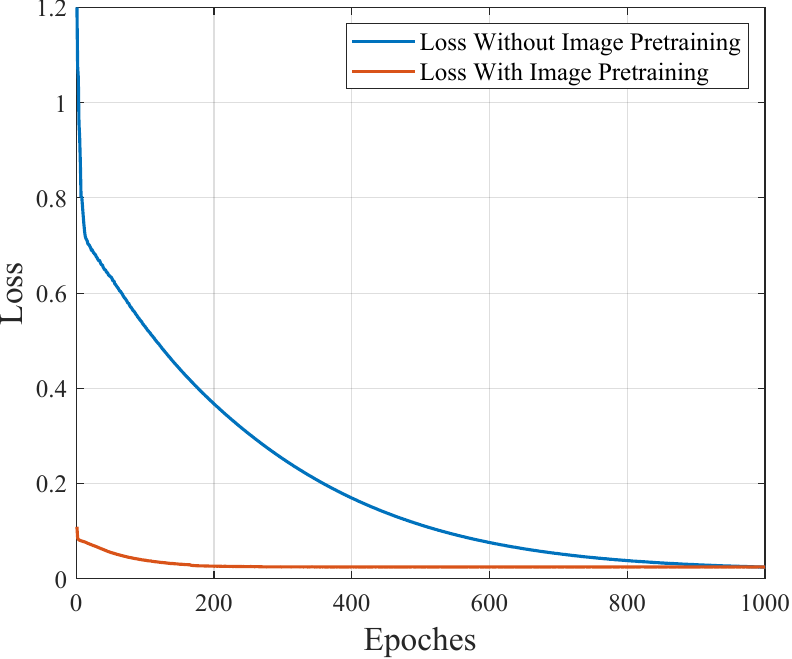}
	\caption{Training loss curves for CSI‑based human pose estimation and localization with and without image pre‑training.}
    \label{fig_IMG2CSI}
\end{figure}

The CDF of the estimation errors of our model on the human pose estimation task is shown in Fig. \ref{CDF}(b). Approximately $50 \%$ of the predictions fall within a $0.2 \text{ m}$ error, while over $80 \%$ lie within $0.247 \text{ m}$. Table~\ref{MAE_joints} presents the mean absolute error (MAE) for each joint, as well as the mean per-joint position error (MPJPE). The calculation method of MPJPE is as follows.
\begin{equation}
    \label{MPJPE}
    \text{MPJPE} = \frac{1}{N}  \sum_{n=1}^{N} \text{MAE}_n,
\end{equation}
where $\text{MAE}_n$ denotes the MAE of the $n$-th joint, and $N$ is the total number of joints. The results show that the overall MPJPE reaches $0.2189\text{ m}$. Our model attains accuracy improvements of $1.5\%$, $11.4\%$, and $5.3\%$ over Protocols 1 to 3 in the MM-Fi baseline, respectively, while using only a quarter of the bandwidth. In Table~\ref{MAE_joints}, the errors for the trunk joints are significantly lower than those for the limb joints. This is attributable, on one hand, to the stronger reflection of Wi-Fi signals from the trunk than from the limbs, and on the other hand, to the strict anatomical constraints on trunk postures, which allow the model to effectively narrow the possible estimation range through prior information. In contrast, the wrists and ankles have higher degrees of freedom, resulting in greater uncertainty.

\begin{figure*}[!t]
    \centering
    \subfloat[]{\includegraphics[width=3.1in]{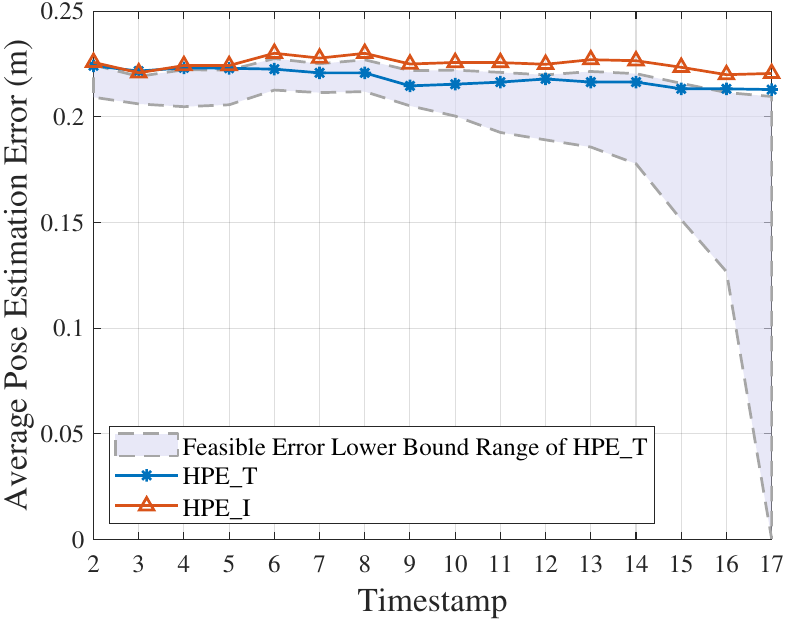}%
        \label{err_HPE_T}}
    \hfil
    \subfloat[]{\includegraphics[width=3.1in]{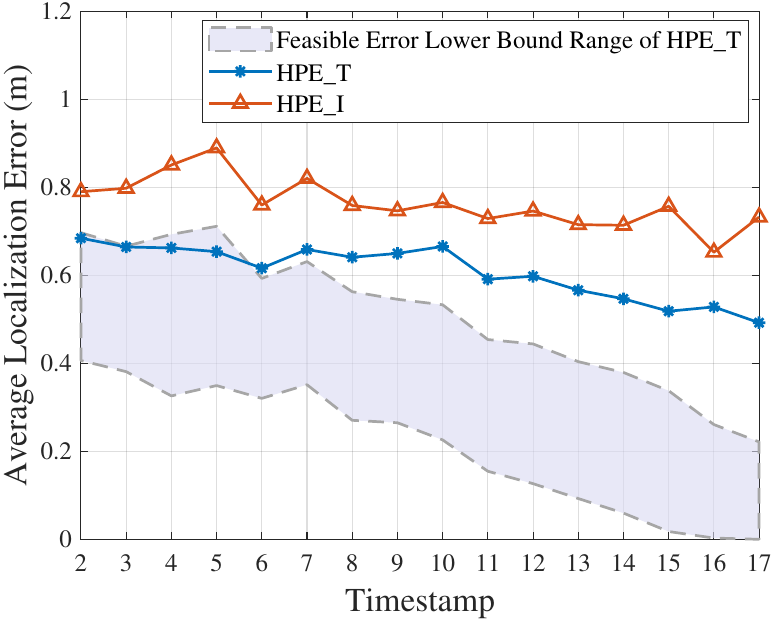}%
        \label{err_locate_T}}
    \caption{Average estimation error and feasible error lower bound range over time. (a) Pose estimation: average errors of HPE\_I and HPE\_T, and the lower bound of HPE\_T. (b) Localization: average errors of HPE\_I and HPE\_T, and the lower bound of HPE\_T.}
    \label{err_t}
\end{figure*} 

We then design an experiment to verify the gain brought by prior information. We compare two training strategies: direct training from scratch, and pre-training on image-based human pose estimation and localization tasks before fine-tuning on CSI data. As shown in Fig. \ref{fig_IMG2CSI}, the pre-initialized model converges significantly faster, requiring roughly half the number of epochs to reach the same loss level as the scratch-trained counterpart. This improvement occurs because the network first learns general structural priors about the human body from image data, and these priors transfer effectively to the CSI modality where the target structure remains the same. This result provides the evidence of prior information gains promised in Section II-B.

To verify the gain brought by temporal correlation, we train the model under two settings: one without temporal memory capability and the other with temporal memory capability, i.e., the full model, resulting in two sets of model parameters, denoted by HPE\_I and HPE\_T, respectively. In addition, to evaluate how far HPE\_T is from its theoretical error lower bound, we use information noise-contrastive estimation (InfoNCE) \cite{oord2018representation} and the leave-one-out (L1Out) upper bound \cite{poole2019variational} to estimate the lower and upper bounds of the conditional mutual information, respectively, and further calculate the feasible region of the theoretical error lower bound of HPE\_T according to Theorem 1, assuming that HPE\_I has already reached its optimum without temporal memory. 

As shown in Fig. \ref{err_t}, HPE\_T consistently outperforms HPE\_I in both localization and pose estimation for most timestamps, and its error decreases over time, confirming the effectiveness of temporal correlation. It is also observed that HPE\_T approaches or even falls below the upper bound of the feasible region of its theoretical error lower bound for the pose estimation task and at timestamps less than 8 for the localization task. Since this region is an estimate of the lower bound rather than the bound itself, this indicates that the estimated region is conservative rather than that the information-theoretic limit has been exceeded. For larger timestamps, however, the performance can be further improved.

Notably, as the timestamp increases, the available batch size decreases due to limited long-sequence data, which expands the estimated feasible error bound range of HPE\_T and shifts both the estimated upper and lower bounds outward. In practice, we only reduce the batch size when the timestamp exceeds 9; therefore, the estimated feasible error lower bound range in Fig. \ref{err_t} remains reliable up to timestamp 9. Beyond this point, however, data sparsity causes the estimated feasible region of the theoretical error lower bound to deviate from the true bound. As a result, for timestamps greater than 9, the actual gap between HPE\_T and the theoretical upper bound is larger than the figure suggests. The same caveat applies to the feasible improvement upper bound range shown in Fig. \ref{err_t}.

\begin{figure}[!ht]
    \centering
	\includegraphics[width=3.1in]{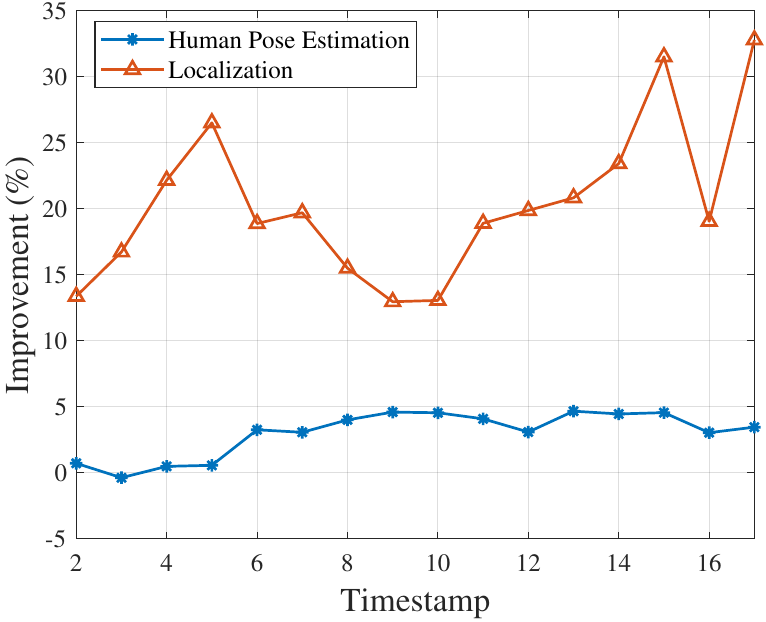}
	\caption{The improvement of HPE\_T over HPE\_I on different timestamps.}
    \label{improvement}
\end{figure}

Fig. \ref{improvement} shows the improvement of HPE\_T over HPE\_I at each timestamp. The improvement of HPE\_T over HPE\_I generally increases with the timestamp. Averaged over all timestamps, temporal correlation reduces the pose estimation error by $2.76\%$ and the localization error by $19.81\%$. This inconsistency is mainly influenced by dimensionality. According to Theorem 1, the relative reduction of the error lower bound is governed by $e^{-\frac{2}{p}I}$ , where p is the dimension of the state to be estimated. The location is a two-dimensional quantity, whereas the pose comprises the three-dimensional coordinates of 16 joints; for a comparable amount of side information, the fractional reduction is therefore considerably smaller for the higher-dimensional state.

\begin{figure}[!ht]
    \centering
	\includegraphics[width=3.2in]{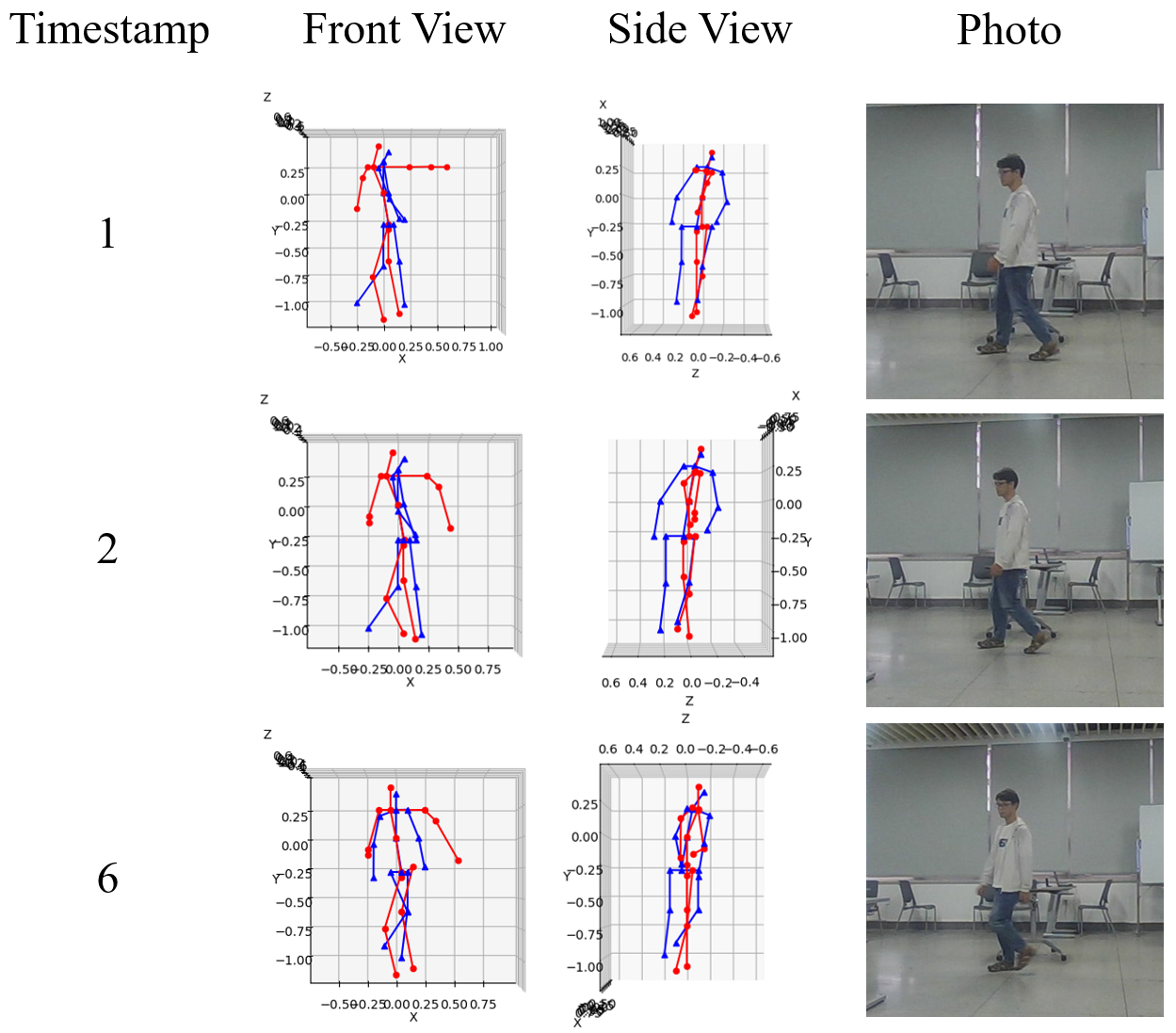}
	\caption{Example of temporal correlation gain, where the blue skeletons represent the ground truth, while the red skeletons indicate the estimation results. It can be observed that the pose estimation results become progressively more accurate over time.}
    \label{example}
\end{figure}

\begin{figure*}[!ht]
    \centering
	\includegraphics[width=6.3in]{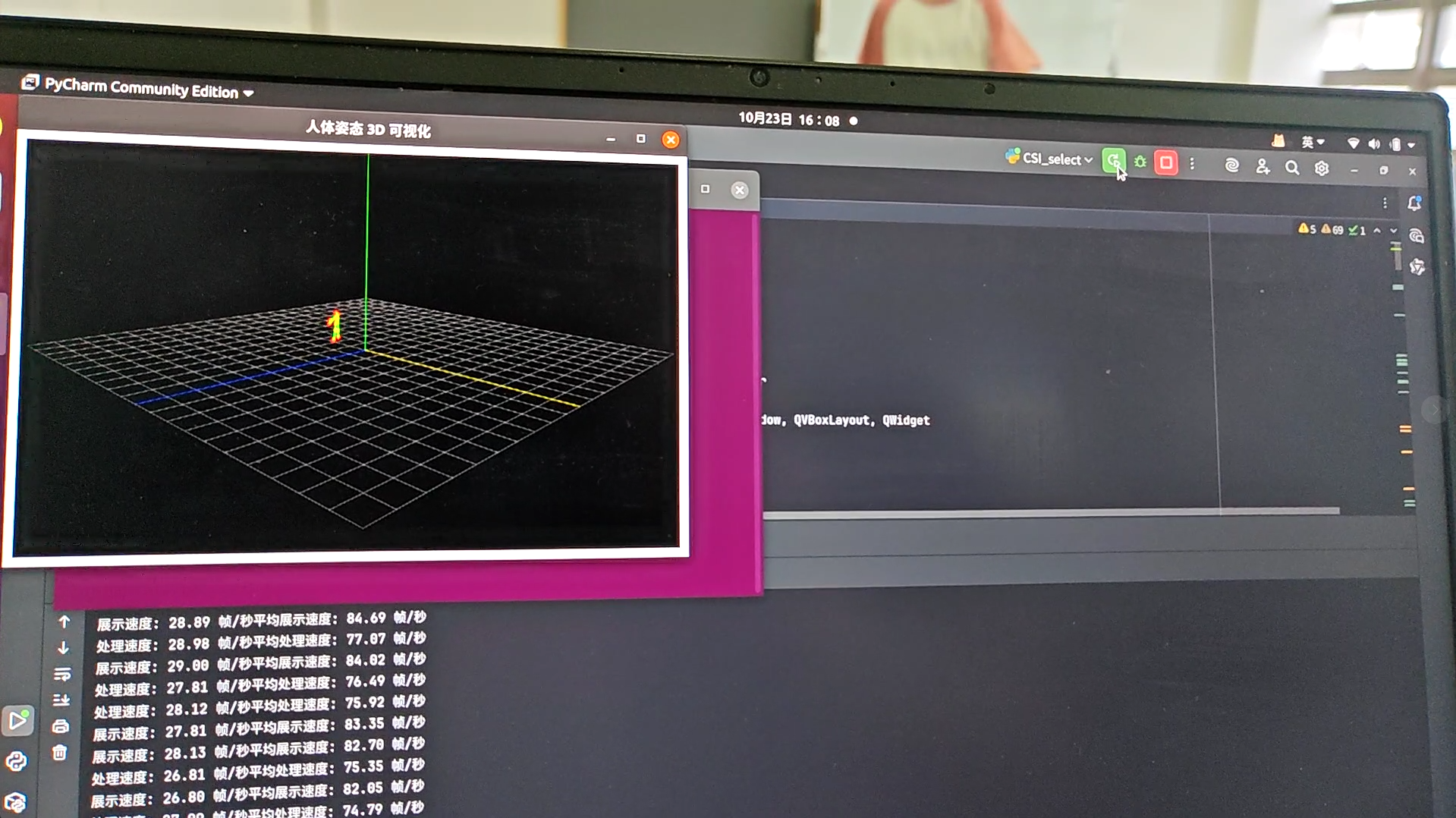}
	\caption{Real-time sensing and visualization system in operation.}
    \label{Real-time_example}
\end{figure*}

We use the example in Fig. \ref{example} to illustrate how temporal correlation reduces estimation errors by narrowing down the number of plausible estimation outcomes. In the figure, the blue skeletons represent the ground truth, while the red skeletons indicate the estimation results. At timestamp 1, the model infers that the subject is raising their hand, and the estimated orientation deviates significantly from the ground truth. By timestamp 2, the estimated pose shows higher similarity to the ground truth, but the orientation remains inaccurate. By timestamp 6, both the pose and orientation converge closely to the ground truth.

A photograph of the real-time sensing and visualization system in operation is shown in Fig. \ref{Real-time_example}. The system achieves an average frame rate of at least $42 \text{ fps}$.

The results above are obtained with a train--test split drawn from the same pool of participants and environments, and therefore characterize the system's accuracy under in-distribution conditions. Two aspects of generalization warrant discussion. With respect to unseen subjects, the structural priors that the network learns are largely subject-independent, whereas the mapping from CSI to pose is shaped by the individual's body dimensions; the former is thus expected to transfer, while the latter would require fine-tuning. With respect to unseen environments, the preprocessing provides a certain degree of invariance, since it explicitly removes the static component, so that the network operates on the dynamic components associated with target motion. Direct transfer to an unseen environment is nevertheless not expected without fine-tuning, because the remaining dynamic components still depend on the geometry of the room.

\section{Conclusions}

This paper theoretically analyzes and identifies two primary sources of performance gains that deep learning brings to Wi-Fi sensing under hardware-constrained conditions. Building on our theoretical findings, we develop a deep learning-based real-time Wi-Fi sensing and visualization system. Within this framework, we show that these gains primarily stem from the effective utilization of prior information and temporal correlation, which act as specific forms of side information that reduce the MSE lower bound. The prior information enables deep learning to generate relatively detailed sensing results based on a vague perception of the target's state by leveraging the structural priors learned during training, without the need to perceive every detail of the target meticulously. This allows deep learning-based sensing to surpass the system's radar aperture. Meanwhile, temporal correlation enables deep learning to reduce the space of plausible sensing outcomes by leveraging the correlation across sequential data, thereby improving sensing accuracy. 

The deep learning-based real-time Wi-Fi sensing and visualization system performs indoor localization and human pose estimation using only a single transceiver pair. The system achieved an average localization error of $0.6124 \text{ m}$, an average pose estimation error of $0.2189 \text{ m}$, and an average frame rate of at least $42 \text{ fps}$ on commodity hardware. Experimental results confirm that temporal correlation directly improves sensing accuracy over time, while prior information is shown to be critical for effective model training. This paper theoretically analyzes the sources of gain in deep learning-based Wi-Fi sensing, validates the analysis through a single-transceiver-pair system, and demonstrates real-time operation. The insights suggest that hardware complexity can be traded for data-driven prior knowledge, pointing toward minimal-hardware sensing systems that achieve high precision through deep learning.

Several aspects remain open for future investigation. The generalization capability of the proposed system and its robustness under low signal-to-noise ratios and substantial occlusion have not been fully investigated, as the present study is confined to a train--test split drawn from the same pool of participants and environments. Extending the framework to these settings, for instance through leave-one-subject-out and cross-environment protocols, and characterizing how the achievable gain varies with the observation quality, are left as future work.

\begin{IEEEbiography}
    [{\includegraphics[width=1in,height=1.25in,clip,keepaspectratio]{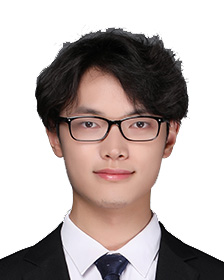}}]
    {Yuxuan Liu} received the B.S. degree in Telecommunication Engineering from Harbin Institute of Technology, Shenzhen, China, in 2022. He is currently pursuing his Ph.D. degree in the School of Electronic and Information Engineering, Harbin Institute of Technology, Shenzhen, China. His research focuses on intelligent signal processing, and integrated sensing and communication.
\end{IEEEbiography}

\begin{IEEEbiography}
    [{\includegraphics[width=1in,height=1.25in,clip,keepaspectratio]{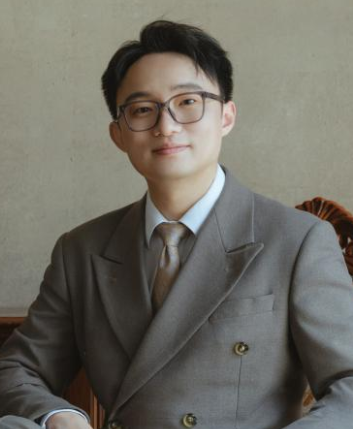}}]
    {Chiya Zhang} (S'15--M'19) received his Ph.D. degree in telecommunication engineering from the University of New South Wales, Sydney, Australia, in 2019. He is currently a Professor at Harbin Institute of Technology, Shenzhen, China. His current research interest is AI applications in telecommunication engineering. He received the Exemplary Reviewer Certificates of the IEEE Wireless Communications Letters in 2018 and IEEE ComSoc Asia-Pacific Outstanding Paper Award in 2020. He is currently serving as an Associate Editor for the IEEE Internet of Things Journal.
\end{IEEEbiography}

\begin{IEEEbiography}
    [{\includegraphics[width=1in,height=1.25in,clip,keepaspectratio]{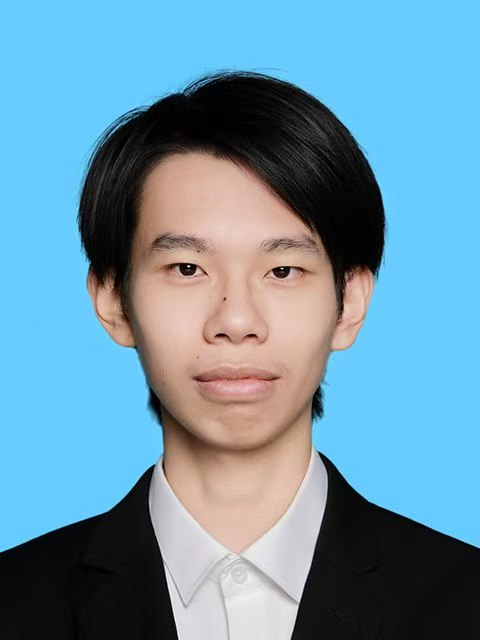}}]
    {Yifeng Yuan} received the B.S. degree in Telecommunication Engineering from Harbin Institute of Technology, Shenzhen, China, in 2024. He is currently pursuing his M.S. degree in the School of Electronic and Information Engineering, Harbin Institute of Technology, Shenzhen, China. His research focuses on intelligent signal processing, and integrated sensing and communication.
\end{IEEEbiography}

\begin{IEEEbiography}
    [{\includegraphics[width=1in,height=1.25in,clip,keepaspectratio]{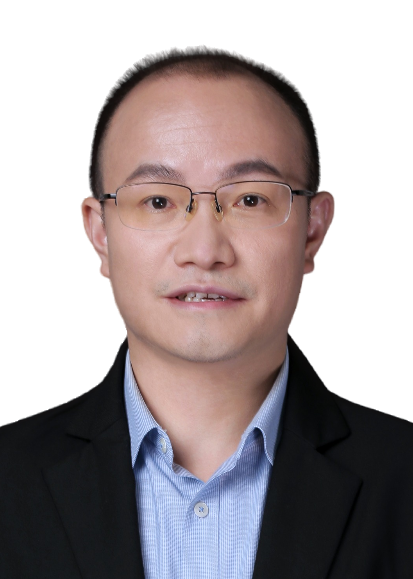}}]
    {Chunlong He} received the M.S. degree in communication and information science from Southwest Jiaotong University, Chengdu, China, in 2010, and the Ph.D. degree from Southeast University, Nanjing, China, in 2014. From September 2012 to September 2014, he was a Visiting Student with the School of Electrical and Computer Engineering, Georgia Institute of Technology, Atlanta, USA. Since 2015, he has been with the College of Electronics and Information Engineering, Shenzhen University, Shenzhen, China, where he is currently an Associate Professor. His research interests include communication and signal processing, green communication systems, channel estimation algorithms, and limited feedback techniques. Dr. He is a member of the Institute of Electronics, Information and Communication Engineers. He is currently an Associate Editor of Digital Communications and Networks.
\end{IEEEbiography}

\begin{IEEEbiography}
    [{\includegraphics[width=1in,height=1.25in,clip,keepaspectratio]{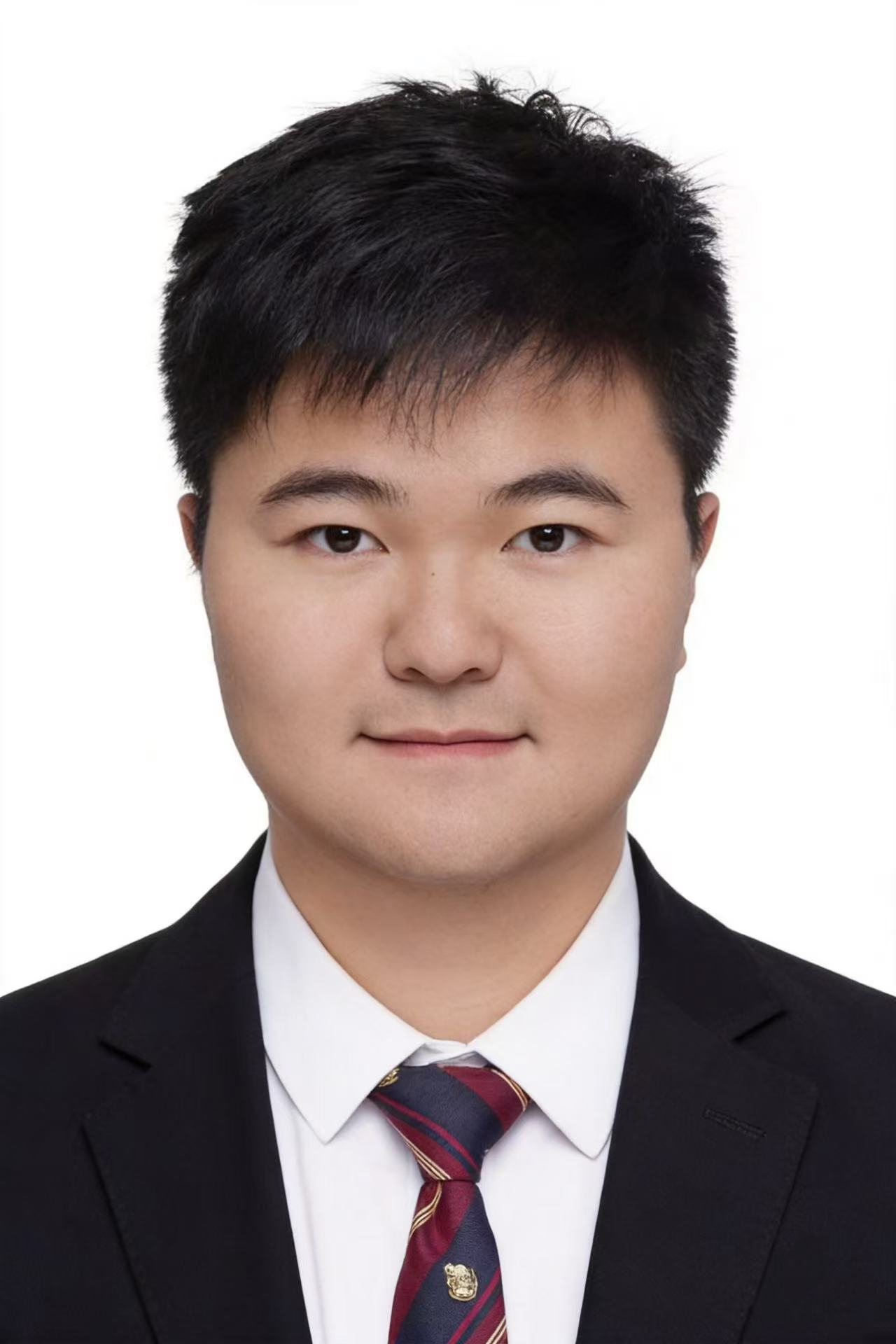}}]
    {Weizheng Zhang} (Member, IEEE) received the B.Eng. degree from Southeast University, China, and the Ph.D. degree from the University of New South Wales (UNSW), Australia, in 2015, and 2019, respectively. From 2020 to 2023, he was a Post-Doctoral Research Fellow at UNSW. He has been a Professor with the School of Information Science and Technology at Harbin Institute of Technology, Shenzhen (HITSZ) since 2024. His current research interests include UAV communications, channel knowledge map, and AI for physical layer communications. He serves as an Editor for IEEE Wireless Communications Letters, Journal of Communications and Information Networks and IEEE Open Journal of Vehicular Technology. He has also served as the TPC co-chair of IEEE/CIC ICCC 2024 and WCSP 2024.
\end{IEEEbiography}

\begin{IEEEbiography}[{\includegraphics[width=1in,height=1.25in,clip,keepaspectratio]{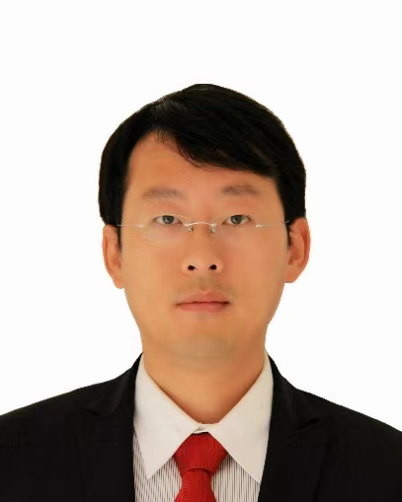}}]{Gaojie Chen}
    (S'09--M'12--SM'18) received the B.Eng. and B.Ec. degrees in electrical information engineering and international economics and trade from Northwest University, China, in 2006, and the M.Sc. (Hons.) and Ph.D. degrees from Loughborough University, U.K., in 2008 and 2012, respectively. After graduation, he took up academic and research positions at DT Mobile, Loughborough University, University of Surrey, University of Oxford and University of Leicester, U.K. He is a Professor and Associate Dean of the School of Flexible Electronics (SoFE) at Sun Yat-sen University, China, and a visiting professor at the 5GIC\&6GIC, University of Surrey, UK. His research interests include wireless communications, flexible electronics, satellite communications, the Internet of Things and secrecy communications. He received the Best Paper Awards from the IEEE IECON 2023 and IEEE GlobeCom 2025, and the Exemplary Reviewer Awards of the IEEE Wireless Communications Letters in 2018, the IEEE Transactions on Communications in 2019 and the IEEE Communications Letters in 2020 and 2021; and Exemplary Editor Awards of the IEEE Communications Letters, IEEE Wireless Communications Letters, and IEEE Transactions on Cognitive Communications and Networking in 2021, 2022, 2023 and 2024, respectively. He served as an Associate Editor for the IEEE Journal on Selected Areas in Communications - Machine Learning in Communications from 2021 to 2022. He serves as an Editor for the IEEE Transactions on Wireless Communications, IEEE Transactions on Cognitive Communications and Networking and IEEE Transactions on Communications, and a Senior Editor for the IEEE Communications Letters and IEEE Wireless Communications Letters, and a Panel Member of the Royal Society's International Exchanges, UK.
\end{IEEEbiography}

\end{document}